\documentclass[12pt,onecolumn,twoside]{IEEEtran}

\usepackage{calc}

\usepackage{multirow}
\usepackage{cite}
\usepackage{graphicx,subfigure}
\usepackage{psfrag}
\usepackage{amsmath,amssymb,amsthm}
\usepackage{color}
\usepackage{tikz} 
\DeclareMathOperator*{\defeq}{\triangleq}

\newtheorem{theorem}{Theorem}
\newtheorem{corollary}{Corollary}[theorem]

\newtheorem{lemma}{Lemma}

\newtheorem{proposition}{Proposition}

\newcommand{\bit}{\begin{itemize}}
\newcommand{\eit}{\end{itemize}}

\newcommand{\bc}{\begin{center}}
\newcommand{\ec}{\end{center}}

\newcommand{\ba}{\begin{array}}
\newcommand{\ea}{\end{array}}

\newcommand{\beq}{\begin{equation}}
\newcommand{\eeq}{\end{equation}}

\newcommand{\beqn}{\begin{equation*}}
\newcommand{\eeqn}{\end{equation*}}

\newcommand{\bean}{\begin{eqnarray*}}
\newcommand{\eean}{\end{eqnarray*}}
\newcommand{\bea}{\begin{eqnarray}}
\newcommand{\eea}{\end{eqnarray}}

\def\hv{\boldsymbol{h}}

\def\sv{\boldsymbol{s}}

\def\xv{\boldsymbol{x}}
\def\yv{\boldsymbol{y}}
\def\zv{\boldsymbol{z}}

\def\Hm{\boldsymbol{H}}
\def\Im{\boldsymbol{I}}

\newcommand{\Dc}{{\mathcal D}}

\newcommand{\Uc}{{\mathcal U}}

\newcommand{\T}{{\scriptscriptstyle\mathsf{T}}}

\newcommand{\CC}{\mathbb{C}}

\renewcommand{\Bmatrix}[1]{\begin{bmatrix}#1\end{bmatrix}}

\newcommand{\dsum}{d_{\text{sum}}}

\newcommand{\TN}{\mathcal{T}_{\text{N}}}
\newcommand{\cond}{\,\vert\,}
\newcommand{\dM}{d_{\text{m}}}
\newcommand{\rM}{R_{\text{m}}}

\begin{document}
\sloppy

\title{Achieving Full DoF in Heterogeneous Parallel Broadcast Channels with Outdated CSIT}
\author{Jinyuan Chen, Sheng Yang, Ayfer \"Ozg\"ur and Andrea Goldsmith
\thanks{Jinyuan Chen, Ayfer \"Ozg\"ur and Andrea Goldsmith are with Stanford University, Department of Electrical Engineering, CA, USA (emails: jinyuanc@stanford.edu, aozgur@stanford.edu, andrea@ee.stanford.edu). Sheng Yang is with the Telecommunications department of SUPELEC, 3 rue Joliot-Curie, 91190 Gif-sur-Yvette, France (email: sheng.yang@supelec.fr). The work was partly supported by the NSF Center for Science of Information (CSoI) under grant NSF-CCF-0939370. The work of Sheng Yang was partly supported by the project FP7 FETOpen HIATUS (grant no. 265578).} 
\thanks{This work was presented in part at ISIT2014.}
}

\maketitle
\pagestyle{headings}

\begin{abstract}


We consider communication over heterogeneous parallel channels, where a transmitter is connected to two users via two parallel channels: a MIMO broadcast channel (BC) and a noiseless rate-limited multicast channel.  We characterize the optimal degrees of freedom (DoF) region of this setting when the transmitter has delayed channel state information (CSIT) regarding the MIMO BC. Our results show that jointly coding over the two channels strictly outperforms simple channel aggregation and can even achieve the instantaneous CSIT performance with completely outdated CSIT on the MIMO BC in the sum DoF sense; this happens when the multicast rate of the second channel is larger than a certain threshold. The main idea is to send information over the MIMO BC at a rate above its capacity and then use the second channel to send additional side information to allow for reliable decoding at both receivers. We call this scheme a two-phase overload-multicast strategy. We show that such a strategy is also sum DoF optimal for  the $K$-user MIMO BC with a parallel multicast channel when the rate of the  multicast channel is high enough and can again achieve the instantaneous CSIT performance (optimal sum DoF) with completely outdated CSIT. For the regime where the capacity of the multicast channel is small, we propose another joint coding strategy which is sum DoF optimal. 
 
\end{abstract}


\section{Introduction}

\begin{figure}[t!]
\centering
\includegraphics[width=11cm]{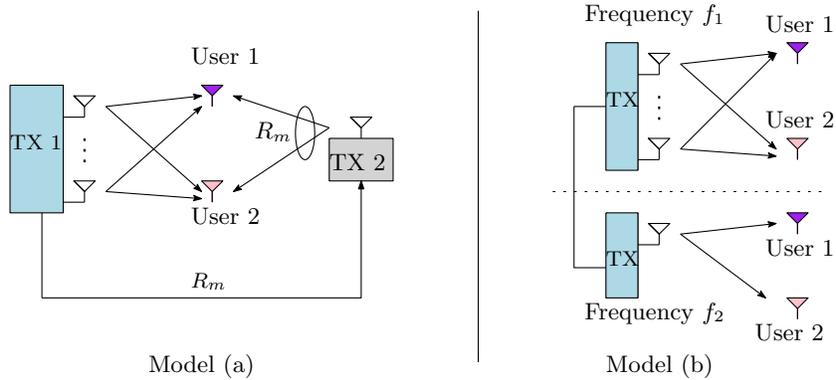}
\caption{(a) The two users can be reached by  transmitter~1 (e.g., base station in a cellular network) directly from the wireless multi-antenna channel and indirectly through the rate-limited backhaul and the local access point~(transmitter~2, e.g., WiFi). 
(b) The transmitter and the users are communicating over two parallel channels, a MISO and a SISO BC.}
\label{fig:LTEWiFi}
\end{figure}

Heterogeneous wireless networks integrate multiple radios with different capabilities, protocol stacks, and spectrum allocations. The flexibility of these different radios allows for more general  dynamic resource allocation, better coverage, and higher capacity.  
In heterogeneous networks, users can be connected to transmitters via parallel channels operating over different networks, such as a cellular and a WiFi network (see Fig.~\ref{fig:LTEWiFi}). In this work we investigate how to optimally communicate over such parallel channels.

We begin with the following setup. A transmitter is connected to two receivers through two parallel channels: the first channel is a multiple-input single-output~(MISO) BC and the second channel is a noiseless rate-limited multicast channel. In a typical realization of our model, the MISO BC (TX~1) can be the cellular downlink from the base station to the mobile users who are also in close proximity to the access
point~(AP) of a WiFi network (TX~2) or a femtocell base station operating over a different frequency (Fig.~\ref{fig:LTEWiFi}~(a)). The base station can therefore transmit to the two users over the cellular downlink while at the same time establish a second multicast channel through the IP network or the femtocell base station.  Alternatively, one can think of the transmitter and the two users as connected by two parallel broadcast channels, a MISO BC operating over one frequency, and a SISO BC operating over another (Fig.~\ref{fig:LTEWiFi}-(b)).

A common and perhaps the simplest way to use the two channels is by \emph{channel aggregation}. That is, the transmitter sends independent information over the two channels and the total throughput becomes the sum
of the individual throughputs of the two channels. This approach typically assumes perfect channel state information at the transmitter (CSIT). However, it is well known that the capacity of the Gaussian multi-antenna BC is very sensitive to the availability of the CSIT. Specifically, in the high SNR regime, the degrees of freedom~(DoF) of the two-user MISO BC are $2$, $4/3$, and $1$ for the cases with instantaneous (perfect) CSIT~\cite{CS:03}, completely outdated CSIT (i.e., the CSIT is available only after the channel's coherence period)~\cite{MAT:11c}, and no CSIT~\cite{HJSV:12}, respectively.
Therefore, given that the first channel is a MISO BC, it is obvious that
channel aggregation will suffer a DoF loss when CSIT is imperfect.

In this work, we propose a simple scheme that strictly outperforms the aforementioned
channel aggregation scheme. We show a somewhat surprising result that, with our scheme, completely outdated CSIT can achieve the same DoF performance as with instantaneous CSIT, provided that the rate of the parallel multicast channel is high enough.  For instance, the proposed scheme can attain a total DoF of $2+\dM$ with completely outdated CSIT when the DoF of the parallel multicast channel~$\dM$ is larger than $2$. This is as if the MISO BC could provide a sum DoF of $2$, which is only possible when instantaneous CSIT is available without the parallel multicast channel.  
The main idea of our optimal scheme (termed a two-phase overload-multicast strategy) is simply to \emph{transmit overload} the MISO BC, i.e, transmit symbols at a rate larger than the multiplexing gain supported by the MISO BC, and then use the multicast channel to \emph{multicast} additional information to enable reliable decoding.
In particular, the additional information sent from the multicast channel is such that it is beneficial for both users but potentially in different ways, e.g.,  it can be used by one user to cancel the created interference  and simultaneously used by the other user as an extra observation for decoding.

Based on this optimal strategy, we characterize the optimal DoF region for a two-user multiple-input multiple-output~(MIMO) BC with a parallel multicast channel. Achieving each corner point in the DoF region involves a careful tuning of the overload and multicast phases of this strategy as well as combining it with zero-forcing and single user transmission. The region is obtained as a function of the multicast channel capacity and the CSIT timeliness  for the MIMO BC. Our result reveals an interesting tradeoff between these two parameters.
Namely, with timely~(e.g., almost instantaneous) CSIT a small multicast channel capacity is
enough to guarantee the maximal sum DoF achievable with instantaneous CSIT, while with completely outdated CSIT a large multicast channel capacity is required to compensate for the sum DoF loss due to the CSIT staleness. In other words, for a given delay of the CSIT
we can determine the  amount of multicast channel capacity needed to achieve the same performance as with instantaneous CSIT; or equivalently, for a given capacity of the multicast channel we can determine the maximal delay we can tolerate in acquiring the CSIT without sacrificing performance. 

Interestingly, this same two-phase overload multicast strategy can be extended to the $K$-user MIMO BC with a parallel multicast channel (in the regime when the number of transmit antennas is larger than the total number of receive antennas in the MIMO BC), and can again achieve the instantaneous CSIT performance (optimal sum DoF) with completely outdated CSIT, provided that the rate of the multicast channel to each user is high enough.
When $K$ is large, the sum DoF gain of the proposed strategy over simple channel aggregation is proportional to the total number of receive antennas. When the capacity of the multicast channel is small, we develop another joint coding strategy which is sum DoF optimal.

The fact that completely stale CSIT can be useful in achieving higher DoF over the $K$-user MISO BC was first revealed by the pioneering work of \cite{MAT:11c}. This work showed the surprising result that completely outdated CSIT achieves a sum DoF performance that surpasses the no-CSIT performance. With $K$-users, their scheme is composed of $K$ phases where in the $k$th phase the transmitter sends so-called $k$-order symbols intended for a group of $k$ users, for $k=1,2,\cdots,K$. Our strategy for the regime where the capacity of the multicast channel is small builds on a similar idea. The MIMO BC transmitter transmits $k$-order symbols in phase $k$ for $k=1,2,\cdots,L$, where $L$, $1\leq L\leq K$, is determined by the available multicasting capacity. The transmission of  higher order symbols is then delegated to the multicast channel. However, this strategy turns out to be suboptimal in the regime where the multicasting capacity is large. The two-phase  overload-multicast strategy that we propose for this case and show to be optimal is strictly different and surprisingly simple. It is composed of only two phases, overload and multicast, with the amount of overloading determined by the number of users and the number of transmit and receive antennas.  

Following \cite{MAT:11c}, the impact of delayed or completely outdated CSIT on the achievable DoF in wireless networks has been studied extensively in the literature (see \cite{YKGY:12d,GJ:12o,CE:13it,TJSP:12,LH:14,CYE:13isit,CE:13MIMO,YYGK:12,VMA:13,VV:12,VV:11t,CEJ:14isit} and also the references therein). 
In all of these works, the performance with \emph{delayed} CSIT turns out to be inferior to the instantaneous CSIT performance with the exception of \cite{LH:14}. This work shows that delayed (but \emph{not completely outdated}) CSIT, with delay less than a $\frac{1}{M+1}$ fraction of the channel coherence time, achieves the same sum DoF as with instantaneous CSIT in the MISO BC setting with $M$ transmit antennas and $M+1$ users.
To the best of our knowledge, whenever there is a DoF performance gap between the cases with instantaneous CSIT and no CSIT, the performance with \emph{completely outdated} CSIT is always inferior to the instantaneous CSIT performance in all of the previous settings considered in the literature (including that of \cite{LH:14}). We believe our work demonstrates the first setting where \emph{completely outdated} CSIT feedback  achieves the same sum DoF as with instantaneous CSIT.

Our work also reveals the value of joint encoding over heterogeneous parallel channels. While
parallel channels have been studied extensively in the literature  \cite{TH:98,Hughes-Hartog:95,T97,LG:01,MZC:06,SCKP:11,SSEP:08,CJ:09}, they usually refer to the realizations of the same physical channel over different time/frequency slots. The parallel channels that emerge in heterogeneous networks which we consider here significantly differ from these earlier models in that the two parallel channels are completely different from each other in nature. While it is known that parallel Gaussian broadcast channels are separable both with single \cite{Hughes-Hartog:95,T97} and multiple antennas \cite{MZC:06}, i.e. the capacity is achieved by using a separate optimized code for each of the channels and then summing up the resultant per-channel capacity, our result reveals that heterogeneous parallel broadcast channels are not always separable (even in the DoF sense).

The remainder of this work is organized as follows. 
Section~\ref{sec:system} describes the system model for the $K$-user MIMO BC with a multicast channel. 
Section~\ref{sec:SchemeExample}  introduces our two-phase overload-multicast strategy via an illustrative example. 
The main results of this work are provided in Section~\ref{sec:mainresult}. 
The achievability and converse  proof details for the two-user MIMO BC with a multicast channel are described in Section~\ref{sec:2user-scheme} and Section~\ref{sec:BC-Outerbound} respectively.
For the $K$-user case, we illustrate our schemes via two examples in Section~\ref{sec:KMIMOexample}, while the general scheme and  the converse  proof are given in the appendices.

Throughout this paper, $(\bullet)^\T$  denotes the transpose operation,  $|\bullet|$ denotes either the magnitude of a scalar or the cardinality of a set. $o(\bullet)$ comes from the standard Landau notation, where $f(x) = o(g(x))$ implies $\lim_{x\to \infty} f(x)/g(x)=0$.
$H(x)$ denotes the entropy of  a random variable $x$, while $h(x)$  denotes the differential entropy of  $x$.
Logarithms are of base~$2$.

\section{System model \label{sec:system} }

\begin{figure}[t!]
\centering
\includegraphics[width=7.2cm]{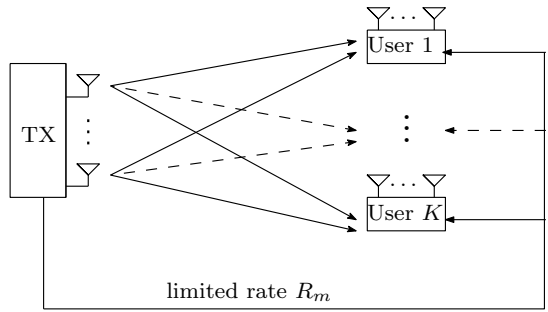}
\caption{$K$-user MIMO BC with a limited-rate  multicast channel.}
\label{fig:BCsidelinkKMIMO}
\end{figure}

\begin{figure}
\centering
\includegraphics[width=10cm]{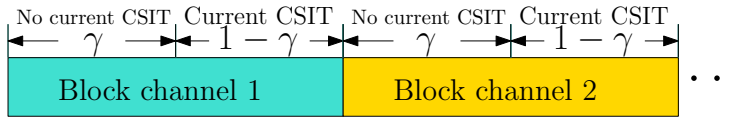}
\caption{Delayed CSIT and block channel model for the MISO BC.}
\label{fig:NotSoDelay}
\end{figure}

We focus on a $K$-user BC in which the transmitter is connected to the receivers through two parallel channels, as shown in Fig.~\ref{fig:BCsidelinkKMIMO}. The first channel is  a $K\times M\times N$  MIMO BC with $M$ $(M\geq KN)$ transmit antennas at the transmitter, and with $N$ receive antennas at each of the $K$ users. The signal vector received over this channel by the $k$th user at time $t$  is given by
\begin{align}
\yv_{k}[t] &=   \Hm_{k}[t] \xv[t]  +   \zv_{k}[t],    \quad   k=1,2,\cdots,K, \label{eq:channelyKMIMO}
\end{align}
where $\Hm_{k}[t]$ denotes the $N\times M$ channel matrix for user~$k$ at time~$t$, $\zv_{k}[t]$ denotes the AWGN noise vector with distribution $\mathcal{CN}(\mathbf{0},\Im)$, and $\xv[t]$ denotes the transmitted signal vector at time $t$ subject to an average power constraint $P$, where $P$ takes the role of the signal-to-noise ratio~(SNR).
We assume a block fading model where  the channel coefficients remain constant during a coherence block of $T_c$ channel uses and change independently from one block to the next. The channel coefficients in each block are independent and identically distributed (i.i.d.) complex Gaussian random variables with distribution $\mathcal{CN}(0,1)$. We assume that the channel coefficients in each block are known to the transmitter only after $\gamma T_c$ channel uses with
$\gamma\in[0,1]$. In other words, during the first $\gamma T_c$ channel uses, the transmitter only knows the channel coefficients corresponding to the past blocks but not the coefficients of the current block. In the remaining $(1-\gamma)T_c$ channel uses, the coefficients of the current block are known perfectly to the transmitter (see Fig.~\ref{fig:NotSoDelay}).  
We call $\gamma$ the CSIT delay fraction hereafter. Note that $\gamma =1$ corresponds to completely outdated CSIT, i.e., the transmitter knows the channel realizations only after the latter have changed to a new independent value.  Throughout this work we assume that all the receivers know all the channel realizations perfectly and instantaneously.
In addition to the connection through the MIMO BC, we assume that the transmitter and the receivers are also connected through a parallel noiseless multicast channel with limited-capacity $\rM$ bits per channel use, over which the transmitter can multicast information to all users. 

In this work, we focus on the high SNR regime and the degrees of freedom performance of the system.  For a given limited rate $\rM$ of the multicast channel, and for an achievable rate tuple $\bigl(R_1,R_2,\cdots, R_K | \rM\bigr)$, where $R_k$ is the rate for user~$k$, the corresponding DoF
tuple $(d_1,d_2,\cdots, d_K)$ is given by 
\begin{align} d_k = \lim_{P \to \infty} \frac{R_k}{\log P}, \quad  k=1,2,\cdots,K .  \label{eq:defd}
\end{align}
 The corresponding DoF region $\Dc$ is then the set of all achievable DoF tuples $(d_1,d_2,\cdots, d_K)$, and the sum DoF  is
\begin{align} \dsum = \sup \bigl\{  d_1+d_2+\cdots+d_K  :  (d_1,d_2,\cdots, d_K) \in \Dc    \bigr\}.  \label{eq:defdsum}
\end{align}
For notational convenience, we assume \[\rM = \dM\log P,\] and, with a slight abuse of terminology,  refer to $\dM$ as the DoF of the multicast channel. $\dM$ measures the multicast channel capacity in $\log P$ units and allows us to relate the multicast channel capacity to the capacity of the MIMO BC in the high SNR limit. Note that with a degrees of freedom approach  we are taking $P$ and therefore the capacity of the MIMO BC to infinity, and therefore we are interested in scaling the capacity of the multicast channel in a comparable way. In the case where the multicast channel is a wireless channel (such as in Fig.~\ref{fig:LTEWiFi}-(b)), $\dM$ corresponds to the DoF of this channel in the classical sense of \eqref{eq:defd}.
Although we focus on the setting with  a \emph{noiseless} multicast channel with  limited-rate $\rM = \dM\log P$,  our results also apply to the setting with an AWGN multicast channel with $\dM$ DoF.

For the simplest case with $N=1$, the first channel becomes a $K$-user MISO BC, and in terms of the notation we will use $y_{k}[t]$, $\hv_{k}[t] \in \CC^{M\times 1}$ and $z_{k}[t]$ to denote the received signal, channel vector, and AWGN noise, respectively, for user~$k$ at time~$t$. From \eqref{eq:channelyKMIMO}, for this special case the channel model is given by:
\begin{align}
y_{k}[t] &=   \hv^{\T}_{k}[t] \xv[t] +   z_{k}[t],    \quad   k=1,2,\cdots, K . \label{eq:channely}
\end{align}

\section{Example of two-phase overload-multicast scheme}\label{sec:SchemeExample}

The design of our scheme depends on the specific parameters of the setting,  as this dictates the optimal use of each one of the parallel channels for purposes such as transmit overloading, side information multicasting, spatial zero forcing, and single user transmission; details of these transmission techniques in our setting will be described below and in Section~\ref{sec:2user-scheme}.  To illustrate the main idea behind the proposed scheme, we start with an example and consider a two-user ($K=2$) MISO BC with $M=2$, $N=1$, $\gamma=1$ and a multicast channel with
$\rM = 2\log P$.
That is, only completely outdated CSI is available at the transmitter. For the sake of simplicity, we let $T_c=1$,
although the result holds for any value of $T_c$.

\begin{figure}
\centering
\includegraphics[width=12cm]{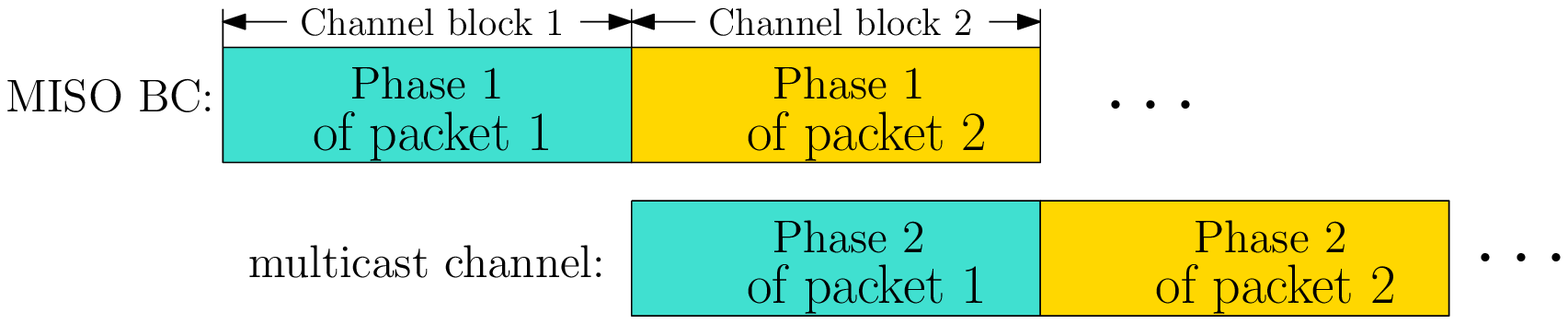}
\caption{Illustration of the two phases for the proposed scheme (with $M=2, N=1, K=2, \dM =2, \ \gamma=1, T_c=1$).
}
\label{fig:BlocksCoop}
\end{figure}

The scheme operates in packets of $2$ symbols
per user. Packet $i$ is communicated over channel use $i$ of the  MISO BC (phase~1) and over channel use $i+1$ of the multicast channel (phase~2), as illustrated in  Fig.~\ref{fig:BlocksCoop}. At the end of these two phases, each receiver can recover its $2$ symbols which yields the optimal $4$ sum DoF for the system. Next, we  describe the transmission in phase~1 and phase~2 for a given packet.

\begin{figure}
\centering
\includegraphics[width=12cm]{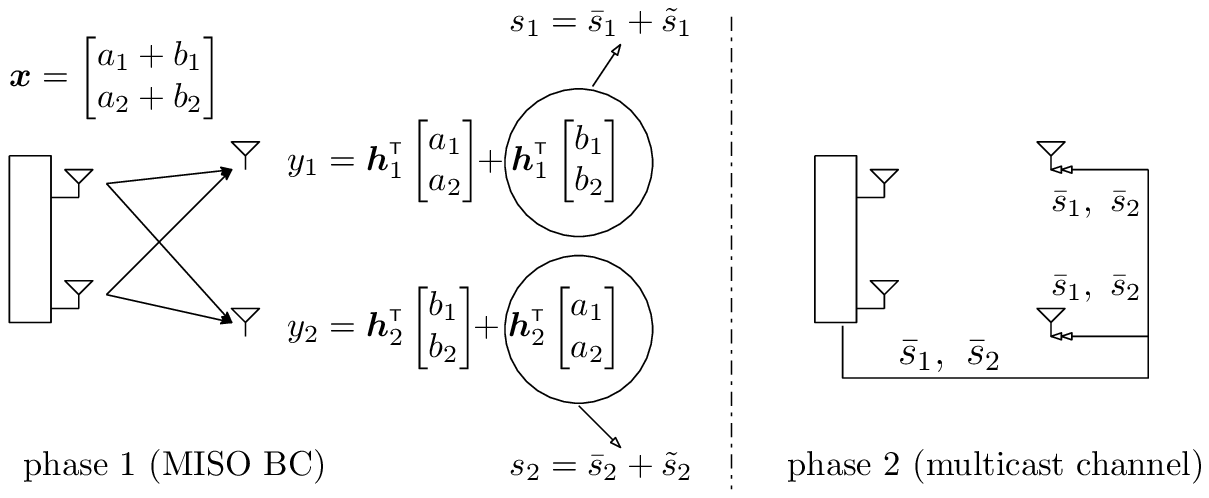}
\caption{Illustration of the two phases for the proposed scheme (with $M=2, N=1, K=2, \dM =2, \ \gamma=1, T_c=1$).}
\label{fig:EffCoop2K}
\end{figure}

\subsubsection{Phase~1 -  transmit overload the MISO BC}

As shown in Fig.~\ref{fig:EffCoop2K}, during phase~1, the transmitter sends four symbols $a_1$, $a_2$,
$b_1$, $b_2$, in one vector in the form (ignoring the time index for simplicity): 
\begin{align} \label{eq:Ph1}
\xv =  \Bmatrix{  a_1 + b_1  \\  a_2 + b_2 },    \end{align}
where symbols $a_1, a_2$ are intended for user 1, symbols $b_1, b_2$ are intended for user 2 and the power of each symbol
is $P/4$.   Then, the received signals at user~1 and user~2 take the form
\begin{align*}
  y_{1} \!=\!   \hv_1^\T \Bmatrix{  a_1  \\  a_2   }   \!+\!   {  \hv_1^\T\Bmatrix{  b_1 \\  b_2   }   }    \!+\!  z_{1},     \quad  \quad 
  y_{2}\!= \!  {  \hv_2^\T\Bmatrix{  a_1  \\ a_2   }   }    \!+  \hv_2^\T \Bmatrix{
  b_1  \\  b_2   }    \! +\!  z_{2}  .
\end{align*}
Note that the total rate of the four symbols \emph{overloaded} as in \eqref{eq:Ph1} surpasses the MISO BC capacity if each symbol carries one DoF.
One can see that, if user~1 is able to learn the two variables $s_{1} \defeq \hv_1^\T\Bmatrix{  b_1 \   b_2   } ^\T
$ and $s_{2} \defeq \hv_2^\T\Bmatrix{  a_1  \    a_2   }^\T  $, then user~1 can remove the interference $s_{1}$ from $y_{1}$, and can use $s_{2}$ as another observation for decoding $a_1$ and $a_2$. 
Similarly, user~2 can decode $b_1$ and $b_2$ with the knowledge of $s_{1}$, $s_{2}$, and $y_2$. Therefore,
in Phase~2, the transmitter will send the information about $s_{1}$ and $s_{2}$ to both users using the
 multicast channel.

\subsubsection{Phase~2 - multicast side information over the parallel multicast channel}
Phase~2 starts after the past CSI about $\hv_1$ and $\hv_2$ is fed back to the transmitter~(see Fig.~\ref{fig:BlocksCoop}). 
The transmitter first \emph{regenerates} $s_{1} $ and $s_{2} $ based on  the past CSI, and then \emph{quantizes}
them into $\bar{s}_{1}$ and $\bar{s}_{2}$ by using $\rM/2$ bits for each. 
Then the transmitter simply sends the total $\rM$ bits of the quantized values
$\bar{s}_{1},\bar{s}_{2}$ to both users through the parallel multicast channel in one channel use (since the multicast channel has capacity $\rM$ bits/channel use).  
After learning $\bar{s}_{1},\bar{s}_{2}$, user~1
and user~2 form their $2\times 2$ MIMO observations of the form
\begin{align*} 
\Bmatrix{ \!y_{1} \!-\!\bar{s}_{1}    \\ \bar{s}_{2}   \! \! }   \!\!\!=\!\!\!  \underbrace{\Bmatrix{  \hv_1^\T
\\ \hv_2^\T}   \! \!  \Bmatrix{ a_1 \\  a_2 } }_{\text{power} \ P}
 \!\!+\! \!\underbrace{\Bmatrix {\!  z_{1}  \!+\! \tilde{s}_{1}   \\      -\tilde{s}_{2} }  \!
 }_{\text{power} \ P^0} ,  \quad \Bmatrix{\! y_{2}\!-\!\bar{s}_{2}    \\ \bar{s}_{1}   \! \! }  \!\! =\!
 \!\underbrace{\Bmatrix{  \hv_2^\T \\ \hv_1^\T}   \!  \! \Bmatrix{ b_1 \\  b_2 }}_{\text{power} \ P}
 \!\!+\! \! \underbrace{\!\Bmatrix { z_{2}  \!+ \! \tilde{s}_{2}   \\
              -\tilde{s}_{1} }\!  \!}_{\text{power} \ P^0}  \ ,
\end{align*}
respectively,  where $\tilde{s}_{1} \defeq  s_{1}- \bar{s}_{1}$ and $\tilde{s}_{2} \defeq s_{2}-\bar{s}_{2}$ are the quantization errors. 
Since the power of $s_1$ and $s_2$ is roughly $P$, it can be easily shown that the variance of the quantization error is roughly
$P 2^{-\frac{\rM}{2}} = 2^{\log P-\frac{\rM}{2}}=1$, i.e., at the noise level. 
Therefore,  with the help of the side information provided from the multicast channel, each user can recover its $2$ symbols from the equivalent $2\times2$ MIMO channel, achieving a sum DoF of $4$ as shown in Fig.~\ref{fig:EffCoop2K}.
A simple cut-set argument reveals that even if instantaneous perfect CSIT were available at the MISO BC transmitter, the performance could not scale better than $4 \log P$ when $P\to\infty$. 
This example shows that completely outdated CSIT can be as good as instantaneous CSIT, in the sum DoF sense.

\section{Main results  \label{sec:mainresult}}

We now return to the general system model of Section~\ref{sec:system} and present the main results of this work. 
Specifically, we first state our result for the  two-user MIMO BC with a multicast channel. Then, we state our results for the $K$-user case.

\subsection{Two-user case \label{sec:2user-result}}

The main result for the two-user MIMO BC with a multicast channel  is summarized in the following theorem.

\vspace{5pt}
\begin{theorem}  \label{thm:2MIMODoFregion}
For the two-user $2\times M \times N$  $(M\geq 2N)$ MIMO BC with a limited-rate  multicast  channel, given the limited rate $\rM=\dM\log P$ and CSIT delay fraction $\gamma$,  the  optimal DoF region is given by
\begin{align}
d_1 &\leq \dM +N,   \label{eq:2Mboundd1} \\
d_2 &\leq \dM +N, \label{eq:2Mboundd2}\\
 d_1+ d_2  &\leq \dM + 2N,  \label{eq:2Mboundd1d2}\\
 2d_1 + d_2 & \leq   2 (\dM + N) +  N(1-  \gamma) ,  \label{eq:2Mbound2d1d2} \\
 2d_2 + d_1  &\leq   2 (\dM + N) +  N(1-  \gamma) .    \label{eq:2Mbound2d2d1}
\end{align}
\end{theorem}
\vspace{5pt}
\begin{proof}
See Section~\ref{sec:2user-scheme} and  Section~\ref{sec:BC-Outerbound} for the achievability and outer bound proofs, respectively.
\end{proof}

\vspace{5pt}

\begin{figure}
\centering
\includegraphics[width=9cm]{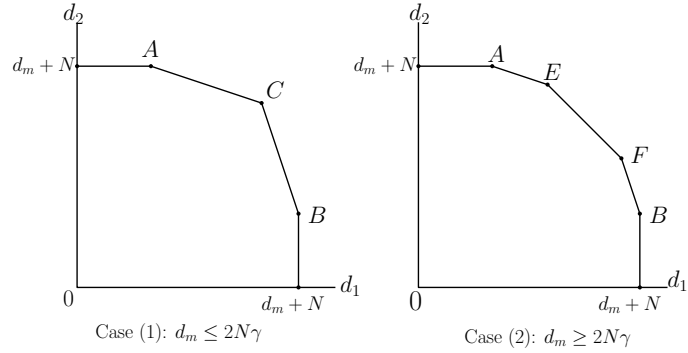}
\caption{Optimal DoF region of the two-user MIMO BC with a parallel multicast channel for the cases with $\dM \leq 2N\gamma$ and $\dM \geq 2N\gamma$. The corner points take the values $A=\bigl(N(1-\gamma), \ \dM+N \bigr)$, $B=\bigl(\dM+N, \ N(1-\gamma) \bigr)$, $E=\bigl( N(1+ \gamma),  \ \dM+N(1-\gamma) \bigr)$, $F=\bigl(\dM+N(1-\gamma), \ N(1+\gamma) \bigr)$, $C=\bigl( \frac{2(\dM+N)+ N(1- \gamma)}{3}, \  \frac{2(\dM+N) + N(1-\gamma)}{3} \bigr)$.}
\label{fig:RegionCoop2KMIMO}
\end{figure}

Fig.~\ref{fig:RegionCoop2KMIMO} depicts the optimal DoF region. Each corner point in the DoF region is achieved by carefully combining the overload-multicast strategy proposed in the previous section with zero-forcing and single user transmission. The following corollary focuses on the sum DoF performance,  which  follows directly from  Theorem~\ref{thm:2MIMODoFregion}.

\vspace{3pt}
\begin{corollary}
  For the two-user $2\times M \times N$  $(M\geq 2N)$ MIMO BC with a limited-rate  multicast  channel, the optimal sum DoF is 
\begin{align}
\dsum  =
\begin{cases}
     2N + \dM   &  \quad \text{if}  \quad   \dM  \geq 2 N \gamma , \\
    \frac{4(\dM + N)  +  2 N(1- \gamma) }{3}   & \quad \text{if}  \quad  \dM  \leq 2 N\gamma .
\end{cases}
\end{align}
\end{corollary}
\vspace{4pt}
 
\begin{figure}
\centering
\includegraphics[width=9cm]{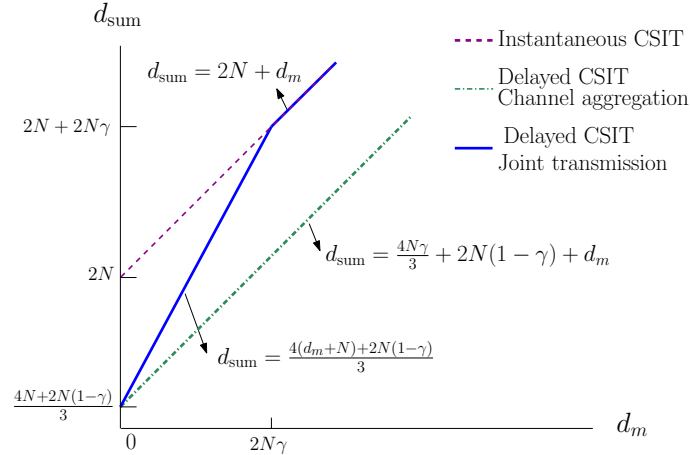}
\caption{Sum DoF $\dsum$ vs $ \dM $ for optimal DoF with full CSIT, optimal DoF with delayed CSIT, and DoF achieved with channel aggregation under delayed CSIT: The two user MIMO BC case.}
\label{fig:SumDoF2MIMOBC}
\end{figure}

For a given CSIT delay fraction $\gamma$, the above result reveals that we need $\rM=2N\gamma\log P$ multicast channel capacity to achieve the instantaneous CSIT performance, in terms of sum DoF. 
As shown in  Fig.~\ref{fig:SumDoF2MIMOBC}, with independent transmissions over the  MIMO BC and parallel multicast channel (channel aggregation) we can only achieve a sum DoF of $4N\gamma/3 +  2N(1- \gamma) + \dM$,
which is strictly suboptimal.

The above result also characterizes the maximum CSIT delay fraction, $\gamma^{*}
\defeq  \arg\max_{\gamma} \{  \dsum (\gamma ) = 2N + \dM\} $, for achieving the maximum sum DoF.

\vspace{3pt}
\begin{corollary}  \label{cor:maxdelay}
For the two-user $2\times M \times N$  $(M\geq 2N)$ MIMO BC with a limited-rate  multicast  channel, the maximum CSIT delay fraction  that achieves the  maximum sum DoF is \[\gamma^{*} = \min\bigl\{ \frac{\dM}{2N},  \  1\bigr\} .\]
\end{corollary}
\vspace{3pt}

For a given multicast channel capacity, Corollary~\ref{cor:maxdelay} gives the maximum delay we can tolerate in feeding back the CSIT without sacrificing perfect CSIT performance, in the sum DoF sense. For example, with $\dM=2N$,  $\gamma^{*}=1$, i.e., completely outdated CSIT is as good as instantaneous CSIT.  On the other hand, with $\dM=N$, we have $\gamma^{*}=1/2$, i.e., we can tolerate a delay of a half coherence period  and still achieve the instantaneous CSIT performance.

\subsection{$K$-user case  \label{sec:KuserMresult}}

In Section~\ref{sec:2user-result}  we have provided the optimal DoF region for the two-user MIMO BC with a multicast channel. Now we move on to the extension to the general $K$-user case ($K\geq 2$), for which we  present sum DoF bounds. For notational convenience, we first define 
 \begin{align}  
  f_p(L, \dM) & \defeq \frac{ K^2\dM +K^2 NL + KNL(K-L)(1-\gamma) }{KL+L(K-L)}  ,  \label{eq:BCKfp}  \\ 
  f_q(L) & \defeq \frac{N\gamma L(L-1) }{ 2K-2L+1}  ,    \label{eq:BCKfq}\\ 
    f_a(L, \dM) & \defeq   \frac{ K(K\!-\!L\!+\!2)(K\!-\!L\!+\!1) \dM  +  KN\gamma (K\!+\!1) (K\!-\!L\!+\!1) }{ (K\!-\!L\!+\!2)L + (K\!+\!1)(K\!-\!L\!+\!1) \sum_{k=1}^{K-L} \frac{1}{k}  }  + KN(1\!-\!\gamma)  ,  \label{eq:BCKfa}  \\ 
  f_b(L) & \defeq    \frac{N\gamma (L-1) }{(K-L+2)\sum_{k=1}^{K-L+1} \frac{1}{k}  } .
 \label{eq:BCKfb}   
  \end{align}

\vspace{5pt}
\begin{proposition}  [Upper bound]\label{pro:KMIMOsum}
For the $K$-user $K\times M \times N$  MIMO BC with a limited-rate  multicast  channel  as described in Section~\ref{sec:system}, the  sum DoF is upper bounded as
\begin{align}   \label{eq:OBkmimobc}
\dsum  \!\leq \!  
\begin{cases}
     \min\Bigl\{ f_p(K, \dM),  \ f_a(1, \dM) \Bigr\}  \! &  \text{if}    \   \    f_q(K)  \leq  \dM  ,  \\
       \min\Bigl\{ f_p(L, \dM),  \  f_a(1, \dM)\Bigr\} \! \! &  \text{if}     \  \  f_q(L)   \leq  \dM \leq f_q(L+1),  \quad \text{for} \quad L=1,2,\cdots, K-1 .
    \end{cases}   
\end{align} 
Note that $\min\Bigl\{ f_p(K, \dM),  \ f_a(1, \dM) \Bigr\} =  f_p(K, \dM) = KN + \dM $ and $\min\Bigl\{ f_p(1, \dM),  \  f_a(1, \dM)\Bigr\} = f_a(1, \dM)=\frac{ K \dM+ KN\gamma}{ \sum_{k=1}^{K}  \frac{1}{ k} } + KN(1-\gamma)$, and $ f_q(K) = K(K-1)N\gamma$.   
\end{proposition}
\vspace{5pt}
\begin{proof}
See Appendix~\ref{sec:OuterbKMIMOBC}.  
\end{proof}
\vspace{5pt}

\begin{proposition} [Lower bound] \label{pro:KMIMOBC}
For the $K$-user $K\times M \times N$ $(M\geq KN)$ MIMO BC with a limited-rate  multicast  channel as described in Section~\ref{sec:system}, the following sum DoF performance is achievable:
\begin{align}  \label{eq:JCkmimobc}
\dsum^{(\text{lb})}  =  
\begin{cases}
     KN + \dM   &  \text{if}   \  \     \dM \geq K(K-1)N\gamma, \\
    \frac{2K}{2K-1} \dM +  KN -  \frac{K(K-1)N\gamma}{2K-1}     &  \text{if}   \ \    f_b ( K  ) \leq \dM  \leq  K(K-1)N\gamma   , \\ 
    f_a ( L, \dM )    &  \text{if}   \ \     f_b ( L )   \leq \dM  \leq  f_b ( L+1 )   \quad \text{for} \quad L=1,2,\cdots, K-1 .
\end{cases}   
\end{align}
Note that $f_b ( 1)=0$ and $ f_b(2) = \frac{ N\gamma}{ K \sum_{k=1}^{K-1}  \frac{1}{ k}} $.
\end{proposition}
\vspace{5pt}
\begin{proof}
See Section~\ref{sec:2user-scheme} for scheme examples and see Appendix~\ref{sec:generalsch} for general proof details.  
\end{proof}
\vspace{5pt}
In the first regime, when $\dM\geq K(K-1)N\gamma$, the lower bound is achieved by an extension of the two-phase overload-multicast strategy which was introduced in Section~\ref{sec:SchemeExample} via a simple example. The last line is achieved by an adaptation of the strategy proposed in \cite{MAT:11c} for the $K$-user MISO BC channel with stale CSIT. Here, the transmission is composed of $K$ phases where 
where in the $k$th phase the transmitter sends so-called $k$-order symbols intended for a group of $k$ users, for $k=1,2,\cdots,K$. We adopt this strategy by performing the first $L$ phases over the MIMO BC and delegating the $(L+1)$th phase to the multicast channel. Here $L$ is chosen carefully between $1$ and $K$ depending on the available multicast capacity. The second line in the proposition is achieved by time sharing between the two strategies corresponding to the first and the third lines. While the DoF region for the two user case in the previous section was achieved by using only the two-phase overload-multicast strategy, specializing the current proposition to $K=2$, one can observe that the optimal sum DoF can be also achieved by the second strategy adopted from \cite{MAT:11c} when $\dM$ is small. However, this strategy fails to achieve the optimal sum DoF when $\dM$ is high (first and second regimes in the proposition) and the two-phase multicast-overload strategy is needed to achieve optimal performance.

Remaining in the same setting of the $K$-user $(M\geq KN)$ MIMO BC with a multicast channel, from the above two theorems we directly get the following conclusion on the optimality of the sum DoF performance.

\vspace{5pt}
\begin{corollary}  [Optimality] \label{cor:Kusers}
The sum DoF lower bound in  \eqref{eq:JCkmimobc}  and upper bound  in \eqref{eq:OBkmimobc} match for the two-user case. For the case with more than two users $(K\geq 3)$, the bounds match when $\dM  \leq  \frac{ N\gamma}{ K \sum_{k=1}^{K-1}  \frac{1}{ k}} $  and  $\dM  \geq  K(K-1)N\gamma$.  
\end{corollary}
\vspace{5pt}

\begin{figure}
\centering
\includegraphics [width=10cm]{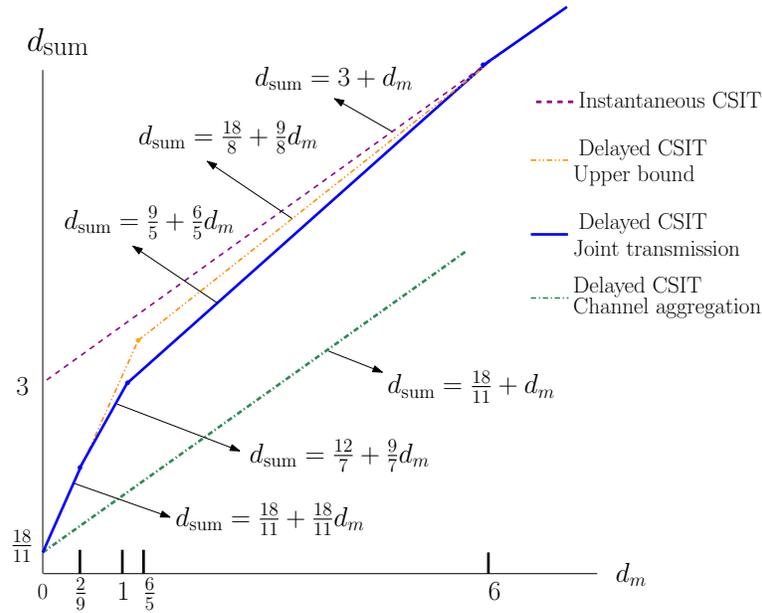}
\caption{Sum DoF $\dsum$ vs $ \dM $ for the $3$-user MISO BC with $ N=1, \gamma=1$ in the following cases: (1) maximal DoF achievable with full CSIT, (2) the DoF upper bound in Proposition~\ref{pro:KMIMOsum}, (3) the DoF achieved in Proposition~\ref{pro:KMIMOBC}, (4) the DoF achieved with channel aggregation. } 
\label{fig:SumDoFK3MISO}
\end{figure}

\noindent Fig.~\ref{fig:SumDoFK3MISO} depicts the sum DoF bounds for the three-user case  with $N=1, \gamma=1$, which are optimal in the regimes of  $\dM  \leq  \frac{ 2}{ 9}$ and of $ \dM  \geq  6$.  

Note that even with instantaneous CSIT $(\gamma =0)$, the sum DoF cannot be larger than $KN+\dM$. Analogous to the 2-user case, the above result shows that delayed CSIT $(\gamma >0)$ achieves the same sum DoF $KN+\dM$ as instantaneous CSIT, provided that $\dM$ is larger than the threshold $K(K-1)N\gamma$. Let $\dM^{*}
\defeq  \arg\min_{\dM} \{  \dsum (\dM) = KN + \dM\} $. 
We have the following corollary.

\vspace{5pt}
\begin{corollary}  [Minimum $\dM$] \label{cor:mindm}
Given the CSIT delay fraction $\gamma$, the minimum value of $\dM$ for achieving the instantaneous CSIT performance (in terms of sum DoF)  is  \[\dM^{*} =K(K-1)N\gamma . \]
\end{corollary}
\vspace{5pt}

For given  DoF $\dM$ for the multicast channel, the above result also characterizes the maximum CSIT delay fraction, $\gamma^{*}
\defeq  \arg\max_{\gamma} \{  \dsum (\gamma ) = KN + \dM\} $, for achieving the instantaneous CSIT performance.

\vspace{3pt}
\begin{corollary}  [Maximum $\gamma$]  \label{cor:maxdelayKuser}
For a given $\dM$, the maximum CSIT delay fraction for achieving the instantaneous CSIT performance (in terms of sum DoF) is \[\gamma^{*} = \min\bigl\{ \frac{\dM}{K(K-1)N},  \  1\bigr\} .\]
\end{corollary}
\vspace{3pt}

For a given multicast channel capacity, Corollary~\ref{cor:maxdelayKuser} gives the delay we can tolerate in feeding back the CSIT without sacrificing perfect CSIT performance, in the sum DoF sense. For example, with $\dM=K(K-1)N$,  $\gamma^{*}=1$, i.e., completely outdated CSIT is as good as instantaneous CSIT.  On the other hand, with $\dM=K(K-1)N/2$,  $\gamma^{*}=1/2$, i.e., we can tolerate a delay of a half coherence period  and still achieve the instantaneous CSIT performance.

Finally, note that with channel aggregation we can only achieve a sum DoF given by
\begin{align}   \label{eq:CAkmimobc}
d_{\text{sum}}^{ (\text{ca})}  = \frac{KN\gamma}{ \sum_{k=1}^{K}  \frac{1}{ k} }  +  KN(1-\gamma) +\dM 
  \end{align}  
(cf.  \cite{MAT:11c, CYE:13isit}).  
Therefore, when $\dM \geq K(K-1)N\gamma$, the sum DoF gain of joint coding (cf. \eqref{eq:JCkmimobc}) over channel aggregation (cf. \eqref{eq:CAkmimobc}) is given by \[  \underbrace{(KN + \dM)}_{\text{joint coding}} - \underbrace{\Bigl( \frac{KN\gamma}{ \sum_{k=1}^{K}  \frac{1}{ k} }  +  KN(1-\gamma) +\dM \Bigr)}_{\text{channel aggregation}}  = KN\gamma \Bigl(1- \frac{ 1}{ \sum_{k=1}^{K}  \frac{1}{ k} } \Bigr)  , \]
which is approximately $KN\gamma$ when $K$ is large.

\section{Achievability for the two-user MIMO BC with a multicast channel \label{sec:2user-scheme}} 

In the illustrative example of Section~\ref{sec:SchemeExample}, the MISO BC was used exclusively for transmit overloading and the parallel multicast channel was used for multicasting side information in order to resolve the resultant interference and provide extra observations for decoding. This was due to the particular choice of $\gamma, \dM, M, N$ and the target DoF point. In this section, we describe the more general scheme for the two-user $M\times N$ $(M\geq 2N)$ MIMO BC with a multicast channel for arbitrary values of $\gamma$ and $\dM$.
Before going into the details, we summarize the following basic strategies and principles for our scheme. 

a) Whenever instantaneous CSIT is available over the MIMO BC, $2N$ fresh symbols are sent with spatial zero-forcing~(ZF) precoding, allowing each user to decode its corresponding $N$ symbols in one channel use. 

b)  When instantaneous CSIT is not available for the MIMO BC, the transmitter can do three different things each for a certain fraction of the coherence block: (i) it can overload the MIMO BC with independent symbols~(over a $\delta $ fraction of the block), (ii) it can multicast side information to enable decoding of symbols overloaded in a previous coherence block~(over a $\theta$ fraction of the block), or (iii) it can  send fresh information to only one of the users (over a $\gamma-\delta -\theta$ fraction of the block since the total fraction for the three possible operations is $\gamma$). 

\begin{figure}
\centering
\includegraphics[width=12cm]{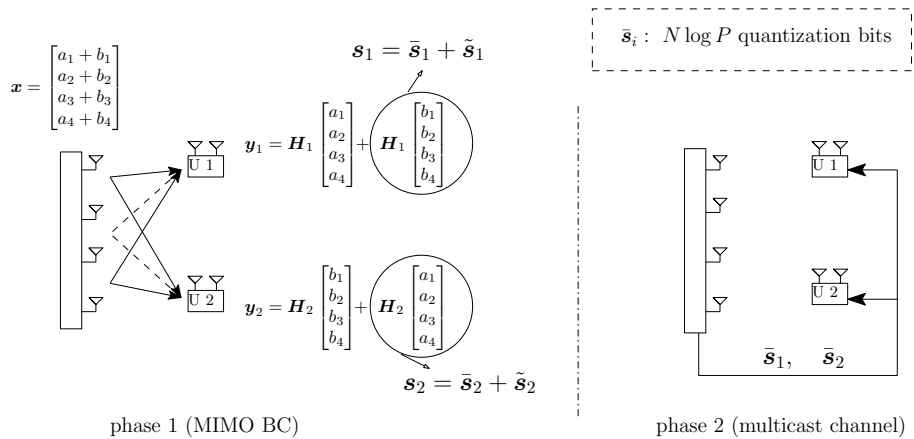}
\caption{Illustration of the two phases for the proposed scheme of a MIMO example (with $K=2, N=2, M=4, \dM =4, \ \gamma=1, T_c=1$). For this example, the optimal sum DoF $2N+\dM$ is achievable with completely outdated CSIT. In the figure, $\bar{\sv}_i$ denotes the quantized version of $\sv_i$ with $2N\log P$ quantization bits, for $i=1,2$. }
\label{fig:BCSLK2N2M4}
\end{figure}

When the transmitter overloads symbols, it transmits $4N$ fresh symbols in one channel use by using the signaling technique in \eqref{eq:Ph1} (also see  Fig.~\ref{fig:BCSLK2N2M4} for an example in the  MIMO case).  In order to decode these $4N$ symbols, we need extra side information of $2N\log P$ bits which should be multicast to both users, as in the illustrative example.

c) The side information generated after the transmit overloading phase is multicast though the multicast channel first and then, if needed, through the MIMO BC. Specifically, the side information is multicast through the MIMO BC as in (b-ii) above only if the multicast channel capacity is insufficient, otherwise it is multicast only through the multicast channel. 

d)  When the multicast channel capacity is very large, using it only a fraction $\eta$ of the time may be sufficient to fully multicast all the generated side information. During the remaining $(1-\eta)$ fraction of time, the multicast channel can be used for  sending  fresh information to one of the users. Note that the multicasting rate and the single user transmission rate over the multicast channel are both $\rM$ bits per channel use. The targeted user depends on the rate pair we would like to achieve. 

\begin{figure}
\centering
\includegraphics[width=10cm]{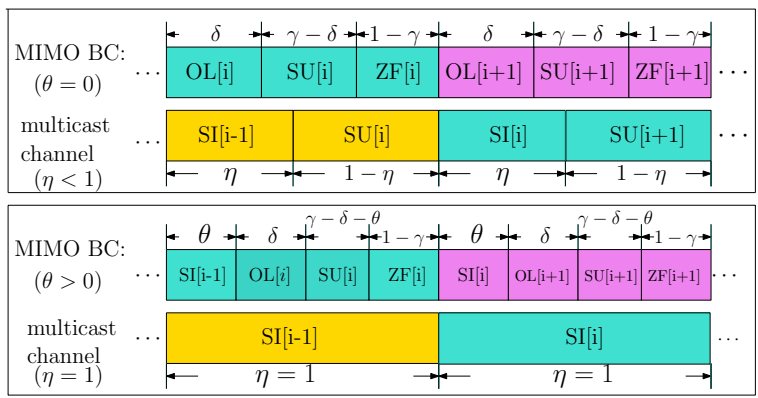}
\caption{Possible operations over the MIMO BC and the multicast channel. The first figure corresponds to the case $\theta=0$ and $\eta \leq 1$  and the second figure corresponds to $\theta>0$ and $\eta=1$. OL, SU, SI, ZF stand for overloading, single user transmission, side information transmission and zero forcing respectively. }
\label{fig:GBlock2MIMO}
\end{figure}

Thus, $\delta, \eta, \theta$ are chosen such that 
    \begin{align}
 2N\delta =  N\theta  +\eta  \dM  \quad  \text{for}  \  0\leq \theta,\delta\leq(\theta \!+\! \delta )\!\leq \!\gamma \leq 1,  \quad  0 \!\leq\! \eta \leq 1  ,  \label{eq:design}
    \end{align}
so that the amount of side information generated for  one block (LHS of \eqref{eq:design}) matches the amount of  side information multicast in the next block (RHS of \eqref{eq:design}). Note that $\theta$ is set to be zero when  $\eta <1$, since the multicast channel  is first used for multicasting side information as stated in c) above (see Fig.~\ref{fig:GBlock2MIMO}).

In both examples with $N=1,  M=2, \dM=2, \gamma=1$ in Section~\ref{sec:SchemeExample} and   with $N=2,  M=4, \dM=4, \gamma=1$ in Fig.~\ref{fig:BCSLK2N2M4},   we chose $\delta=\gamma,   \eta =1, \theta =0$ for achieving the optimal sum DoF. 
In the following, we provide the explicit values of $(\delta, \eta, \theta)$ for achieving each corner point of the DoF region in Fig.~\ref{fig:RegionCoop2KMIMO} of the general setting. Note that  time sharing between these corner points gives the full DoF region.

\subsection{Corner points $A$ and $B$}
The corner points $A=\bigl( N(1-\gamma), N+\dM\bigr)$ and $B=\bigl(  N+\dM, N(1-\gamma) \bigr)$ are achievable
for any values of $\dM$ and $\gamma$ by a simple channel aggregation strategy, which is equivalent to the general scheme by fixing  $\delta=\theta =\eta=0$. 
Point~$A$ is achieved when only user~$2$ is chosen for single user transmission over both the BC and the multicast channel, i.e.,
\begin{align*}
 d_1=   \underbrace{N(1\!-\! \gamma ) }_{\text{ZF}},   \quad \quad    d_2 = \!\!\! \underbrace{\dM }_{\text{single user}} \!\!+\!\!\underbrace{N\gamma }_{\text{single user}} \!\!+  \!\underbrace{N(1\!-\! \gamma ) }_{\text{ZF}} .
\end{align*}
Note that single user transmission over the multicast channel provides $\dM$ DoF, single user transmission over the MIMO BC provides $N$ DoF, and zero forcing provides $N$ DoF per user.  The last two operations are performed in $\gamma$ and $1-\gamma$ fractions of the time respectively. 
Similarly, point~$B$ is achieved when only user~$1$ is chosen for single user transmission.

\subsection{Corner point $C$ when $\dM \le 2 N\gamma$}
To achieve point $C=\bigl( \frac{2(\dM+N)+ N(1- \gamma)}{3}, \  \frac{2(\dM+N) + N(1-\gamma)}{3} \bigr)$, we use the general scheme by setting $\delta= \frac{N\gamma + \dM}{3N}$,   $\theta = \frac{2N\gamma  - \dM}{3N}$,  $\eta = 1$  (cf. \eqref{eq:design}). Note that since $\theta+\delta=\gamma$ and $\eta = 1$, there is no single user transmission in this case. Since the multicast channel capacity is insufficient ($\dM \le 2 N\gamma$), in addition to the multicast channel, the MIMO BC is used in some fraction of time for multicasting the side information, which allows us to achieve:
\begin{align*}
  d_1 \!=\! d_2 \!=  \! \! \! \underbrace{ 2 N\delta }_{\text{overloading}} \! \!  \!+ \!    \underbrace{ N(1\!-\! \gamma ) }_{\text{ZF}}=\frac{2(\dM+N)+ N(1- \gamma)}{3} .
\end{align*}
Note that during overloading we transmit $2N$ symbols per user, hence achieve $2N$ DoF per user once the interference is resolved and the extra observation is obtained.
Since side information multicasting does not provide any fresh information it does not contribute to the DoF computation given above.

\subsection{Corner points $E$ and $F$ when $\dM \ge 2 N\gamma$}
In this case, we use the general scheme by setting $\delta=\gamma$,   $\eta =2N\gamma/\dM$, $\theta =0$. 
With $\dM \ge 2 N\gamma$, now the multicast channel is used partially for multicasting the side information and partially for single user transmission, which yields the following sum DoF 
\begin{align*}
  d_1 + d_2= \underbrace{4N\delta}_{\text{overloading}}  + \underbrace{(1-\eta)\dM }_{\text{single user}} +    \underbrace{2N(1\!-\! \gamma ) }_{\text{ZF}}  = 2N+  \dM .
\end{align*}
As a result, point~$E=\bigl( N(1+ \gamma),  \ \dM+N(1-\gamma) \bigr)$  is achieved when only user~$2$ is chosen for single user transmission, while point~$F=\bigl(\dM+N(1-\gamma), \ N(1+\gamma) \bigr)$ is achieved when only user~$1$ is chosen for single user transmission.

\section{Converse for the two-user MIMO BC with a multicast channel\label{sec:BC-Outerbound}} 

In this section we provide the converse proof for the two-user MIMO BC with a multicast channel (cf. Theorem~\ref{thm:2MIMODoFregion}).  
Essentially, the proof is based on Fano's inequality, basic entropy inequalities, genie-aided  techniques, as well as the symmetric entropy technique. 
We let $\yv^{n}_k$  denote the signals received from the MIMO BC over  $n$ consecutive channel uses by receiver~$k$,   $y^{n}_0$ denote the multicast channel outputs,   $W_k$  denote the message intended for receiver~$k$, and $\Omega^{n}$ denote the set of all channel states.  

At first we provide the following lemma to be used. The proof of this lemma uses the symmetric entropy technique.  
\vspace{5pt}
\begin{lemma}   [Symmetric entropy] \label{lm:DiffEntroMIMOBC} 
 $h(\yv^{n}_2, \yv^{n}_1 |W_1,\Omega^{n})    -  2 h(\yv^{n}_1|W_1, \Omega^{n})   \leq  nN(1- \gamma )
\log P      +    n\cdot o(\log P).     $
\end{lemma}
\vspace{5pt}
\begin{proof}
Let $\TN \defeq \left\{ t\in[1,n]: \, \text{the current channel state
is not known at time $t$} \right\}$. Note that $|\TN|=n\gamma$, i.e., the total time of communication without current CSIT is $n\gamma$.  Then, we have
\begin{align} 
&h(\yv^{n}_2, \yv^{n}_1 |W_1,\Omega^{n})    -  2 h(\yv^{n}_1|W_1, \Omega^{n})   \nonumber\\
&= \!\sum_{t=1}^n \bigl(h(\yv_{2}[t], \yv_{1}[t] |\yv^{t-1}_{2}, \yv^{t-1}_{1},W_1,\Omega^{n})   \! - \!  2 h(\yv_{1}[t]|\yv^{t-1}_1,W_1, \Omega^{n}) \bigr)   \nonumber \\
&\leq \sum_{t=1}^n \bigl(h(\yv_{2}[t], \yv_{1}[t] |\yv^{t-1}_{1},W_1,\Omega^{n})
-  2 h(\yv_{1}[t]|\yv^{t-1}_1,W_1, \Omega^{n}) \bigr)   \label{eq:Mcondentro598}
 \\
&\leq \sum_{t=1}^n \bigl(h(\yv_{2}[t] |\yv^{t-1}_{1},W_1,\Omega^{n}) - h(\yv_{1}[t]|\yv^{t-1}_1,W_1, \Omega^{n}) \bigr)   \label{eq:Mcondentro599}  \\
&= \sum_{t\not\in \TN} \bigl(h(\yv_{2}[t] |\yv^{t-1}_{1},W_1,\Omega^{n}) -
h(\yv_{1}[t]|\yv^{t-1}_1,W_1, \Omega^{n}) \bigr)   \label{eq:Mcondentro597}  \\
&\le  (n - |\TN|) \bigl(N\log P+o(\log P) \bigr)  ,    \label{eq:Mcondentro432}
\end{align}
where \eqref{eq:Mcondentro598} uses  the fact that conditioning reduces differential
entropies;  \eqref{eq:Mcondentro599} follows from $h(\yv_{2}[t], \yv_{1}[t]
|\yv^{t-1}_{1},W_1,\Omega^{n}) \le h(\yv_{1}[t] |\yv^{t-1}_{1},W_1,\Omega^{n}) +
h(\yv_{2}[t] |\yv^{t-1}_{1},W_1,\Omega^{n})$; \eqref{eq:Mcondentro597} is due to the
symmetry of the output whenever the channel input is independent of the current
channel state, i.e., $ h(\yv_{1}[t] |\yv^{t-1}_{1},W_1,\Omega^{n}) =  h(\yv_{2}[t]
|\yv^{t-1}_{1},W_1,\Omega^{n})$ whenever $t\in\TN$; and the last inequality holds since
$h(\yv_{2}[t] | \yv^{t-1}_{1},W_1,\Omega^{n})\le N\log P+o(\log P)$ and 
$h(\yv_{1}[t] | \yv^{t-1}_{1},W_1,\Omega^{n})\ge h(\yv_{1}[t] | \yv^{t-1}_{1},W_1,W_2,\Omega^{n}) = h(\zv_{1}[t]) = o(\log P)$. 
Finally, by subsisting  $|\TN|$ with $n\gamma$, we complete the proof. 
\end{proof}

Now we first prove the outer bound corresponding to \eqref{eq:2Mboundd1}. Starting from Fano's inequality, we have
\begin{align}
 nR_1&\leq I(W_1; \yv^{n}_1, y^{n}_0 |\Omega^{n})  +  n \epsilon_n  \nonumber\\
&=  I(W_1; \yv^{n}_1 |\Omega^{n})  + I(W_1; y^{n}_0 |\yv^{n}_1, \Omega^{n})  + n \epsilon_n\\
&= h(\yv^{n}_1 | \Omega^n) - h(\yv^{n}_1 | W_1, \Omega^n) + H(y^{n}_0 |\yv^{n}_1, \Omega^{n}) - H(y^{n}_0 |W_1, \yv^{n}_1, \Omega^{n}) +  n \epsilon_n \label{eq:2Mtmp781}\\
&\le nN\log P + n\rM - h(\yv^{n}_1 | W_1, \Omega^n)   - H(y^{n}_0 |W_1, \yv^{n}_1, \Omega^{n}) +  n\cdot o(\log P) \label{eq:2Mtmp785}\\
&\le nN\log P + n \rM + n\cdot o(\log P) , \label{eq:2Mtmp786}
\end{align}
where \eqref{eq:2Mtmp785} follows from
$h(\yv^n_1 | \Omega^n)\le nN \log P + n\cdot o(\log P)$ and the rate constraint of the multicast
channel $H(y^{n}_0 |\yv^{n}_1, \Omega^{n}) \le H(y^{n}_0) \le n \rM$; the last inequality follows from
the non-negativity of the entropy  $ H(y^{n}_0 |W_1, \yv^{n}_1, \Omega^{n})$ and the fact that
$h(\yv^n_1 | W_1, \Omega^n) \ge h(\yv^n_1 | W_1, \xv^n, \Omega^n) = h(\zv^n_1) = n\cdot o(\log P)$.
Hence, dividing \eqref{eq:2Mtmp786} by $n \log P$ and let $P\to\infty$, \eqref{eq:2Mboundd1}
follows immediately and so does \eqref{eq:2Mboundd2} due to the symmetry.

Following similar steps as above, \eqref{eq:2Mboundd1d2} can also be derived as:  
\begin{align}
& nR_1+nR_2 \nonumber\\
 &\leq I(W_1,W_2; \yv^{n}_1,  \yv^{n}_2, y^{n}_0 |\Omega^{n})  +  n \epsilon_n  \nonumber\\
&=  I(W_1,W_2; \yv^{n}_1,  \yv^{n}_2 |\Omega^{n})  + I(W_1,W_2; y^{n}_0 |\yv^{n}_1, \yv^{n}_2, \Omega^{n})  + n \epsilon_n \nonumber\\
&= h(\yv^{n}_1,  \yv^{n}_2 | \Omega^n) - h(\yv^{n}_1, \yv^{n}_2 | W_1,W_2, \Omega^n) \nonumber\\ & \quad + H(y^{n}_0 |\yv^{n}_1, \yv^{n}_2, \Omega^{n}) - H(y^{n}_0 |W_1,W_2, \yv^{n}_1, \yv^{n}_2, \Omega^{n}) +  n \epsilon_n \nonumber\\
&\le 2Nn\log P + n\rM - h(\yv^{n}_1, \yv^{n}_2 | W_1,W_2, \Omega^n)  - H(y^{n}_0 |W_1,W_2, \yv^{n}_1, \yv^{n}_2, \Omega^{n}) +  n\cdot o(\log P) \label{eq:2Mtmp324}\\
&\le 2Nn\log P + n \rM + n\cdot o(\log P) , \label{eq:2Mtmp432}
\end{align}
where \eqref{eq:2Mtmp324} follows from
$h(\yv^{n}_1, \yv^{n}_2 | \Omega^n)\le 2Nn \log P + n\cdot o(\log P)$ and the rate constraint of the multicast channel $H(y^{n}_0 |\yv^{n}_1, \yv^{n}_2 , \Omega^{n}) \le n \rM$; the last inequality follows from
the non-negativity of the entropy  $ H(y^{n}_0 |W_1,W_2, \yv^{n}_1, \yv^{n}_2 , \Omega^{n})$ and the fact that
$h(\yv^{n}_1, \yv^{n}_2  | W_1,W_2, \Omega^n) \ge h(\yv^{n}_1, \yv^{n}_2  | W_1,W_2, \xv^n, \Omega^n) = h(\zv^n_1,\zv^n_2) = n\cdot o(\log P)$. At this point, \eqref{eq:2Mboundd1d2}
follows easily.

We proceed to prove the outer bound \eqref{eq:2Mbound2d1d2}. Giving the side information $\{ \yv^{n}_1,W_1\}$ to user~2, we obtain 
\begin{align}
 &nR_2  \nonumber \\ 
 &\leq I(W_2; W_1,\yv^{n}_2, \yv^{n}_1, y^{n}_0|\Omega^{n})  +  n \epsilon_n  \nonumber \\
 &= I(W_2; W_1|\Omega^{n}) +I(W_2; \yv^{n}_2, \yv^{n}_1, y^{n}_0|W_1,\Omega^{n})  +  n \epsilon_n  \nonumber \\
&= I(W_2; \yv^{n}_2, \yv^{n}_1 |W_1,\Omega^{n}) +I(W_2; y^{n}_0 |\yv^{n}_2, \yv^{n}_1,
W_1,\Omega^{n})   +  n \epsilon_n \label{eq:tmp98273}  \\
&= \! h( \yv^{n}_2, \yv^{n}_1 |W_1,\Omega^{n}) \!-\! h( \yv^{n}_2, \yv^{n}_1 |W_1,W_2,\Omega^{n}) \!+\!  H ( y^{n}_0 |\yv^{n}_2, \yv^{n}_1, W_1,\Omega^{n}) \!-\! H ( y^{n}_0 |\yv^{n}_2, \yv^{n}_1,W_1, W_2, \Omega^{n})  \!+\!  n \epsilon_n \nonumber \\
&\le h(\yv^{n}_2, \yv^{n}_1 |W_1,\Omega^{n}) + H(y^{n}_0 |\yv^{n}_2, \yv^{n}_1, W_1,\Omega^{n}) + n\cdot o(\log P)  ,\label{eq:2Mtmp782}
\end{align}
where \eqref{eq:tmp98273} uses the independence between $W_1$ and $W_2$, the last inequality follows from $h(\yv^n_1, \yv^n_2 | W_1, W_2, \Omega^n) \ge n\cdot o(\log P)$ and the non-negativity property of the entropy. 
Finally, combining \eqref{eq:2Mtmp785} and \eqref{eq:2Mtmp782}, we get 
\begin{align}
  \lefteqn{n (2 R_1 + R_2)} \nonumber\\
&\leq 2 n N\log P + 2n \rM  + H(y^{n}_0 |\yv^{n}_2, \yv^{n}_1, W_1,\Omega^{n}) - 2 H(y^{n}_0 |W_1, \yv^{n}_1, \Omega^{n}) \nonumber \\ 
& \quad   + h(\yv^{n}_2, \yv^{n}_1 |W_1,\Omega^{n}) - 2 h(\yv^n_1 | W_1, \Omega^n) + n\cdot o(\log P) \nonumber \\
&\le 2 n N \log P + 2 n \rM - H(y^{n}_0 |W_1, \yv^{n}_1, \Omega^{n}) \nonumber \\
&\quad + h(\yv^{n}_2, \yv^{n}_1 |W_1,\Omega^{n}) - 2 h(\yv^n_1 | W_1, \Omega^n) + n\cdot o(\log P)
\label{eq:2Mtmp893} \\
&\le 2 n N \log P + 2 n \rM + nN(1- \gamma ) \log P  + n\cdot o(\log P) , \label{eq:2Mtmp932}
\end{align}
where \eqref{eq:2Mtmp893} follows from the fact that removing conditions increases entropy; 
the last inequality follows from the non-negativity of entropy $H(y^{n}_0 |W_1, \yv^{n}_1,
\Omega^{n})$ and Lemma~\ref{lm:DiffEntroMIMOBC}. 
Dividing \eqref{eq:2Mtmp932} by $n \log P$ and let $P\to\infty$, we can
obtain \eqref{eq:2Mbound2d1d2}, and then \eqref{eq:2Mbound2d2d1} by the symmetry of the setting.

\section{Scheme examples for the $K$-user case} \label{sec:KMIMOexample}  

In this section we illustrate our schemes for the $K$-user $K\times M \times N$ $(M\geq KN)$ MIMO BC with a multicast channel via two examples. Our first scheme for the $K$-user case is an extension of the two-phase overload-multicast strategy we have illustrated for the 2-user case in Section~\ref{sec:2user-scheme}.  The main idea is again to \emph{transmit overload} the MIMO BC, i.e, transmit symbols at a rate larger than the multiplexing gain supported by the MIMO BC, and then use the multicast channel to \emph{multicast} additional information to enable reliable decoding. The number of symbols transmit overloaded over the MIMO BC is given by $K^2N$. This strategy turns out to be optimal in the regime where $\dM \geq K(K-1)N\gamma$). For the case when $\dM$ is smaller than this threshold, we present another strategy which has a similar two-phase flavor but builds on the scheme proposed in \cite{MAT:11c}.

To illustrate the proposed schemes, we here provide two examples, one for the case with large and the other  with small $\dM$ respectively. The general scheme and the outer bound proof are given in the appendices.

\subsection{Illustrative example $(K=3, N=2, M=6, \dM =12, \ \gamma=1)$}\label{sec:SchemeExampleKMIMO}

We first consider the example with $K=3, N=2, M=6, \rM=12\log P, \ \gamma=1$ (completely outdated CSI).
Again we let $T_c=1$ for the sake of simplicity.

Our scheme operates in packets of $6$ symbols
per user. Similarly to the previous case (cf. Section~\ref{sec:SchemeExample}), packet $t$ is communicated over channel use $t$ of the  MIMO BC (phase~1) and channel use $t+1$ of the multicast channel (phase~2), as shown in Fig.~\ref{fig:BlocksCoop}. At the end of these two phases, each receiver can recover its $6$ symbols which yields the optimal $18$ sum DoF for the system. Next, we  describe the transmission in phase~1 and phase~2 for a given packet.

\begin{figure}
\centering
\includegraphics[width=14cm]{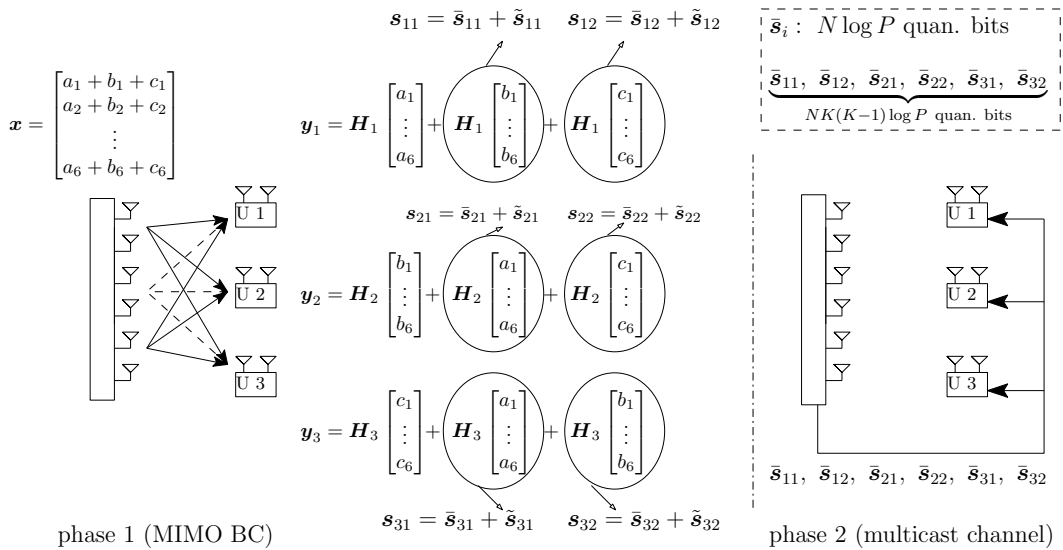}
\caption{Illustration of the two phases for the proposed scheme (with $K=3, N=2, M=6, \dM =12, \ \gamma=1, T_c=1$). For this example, the optimal sum DoF $KN+\dM$ is achievable with completely outdated CSIT.}
\label{fig:BCSLK3N2M6}
\end{figure}

\subsubsection{Phase~1 -  transmit overload the MIMO BC}

As shown in Fig.~\ref{fig:BCSLK3N2M6}, during phase~1, the transmitter sends 18 symbols $\{a_i, b_i, c_i\}_{i=1}^{6}$, in one vector of the form (ignoring the time index for simplicity): 
\begin{align} \label{eq:Ph1KMIMO}
\xv =  \Bmatrix{  a_1 + b_1 +c_1 \\  a_2 + b_2+c_2 \\ \vdots  \\  a_6 + b_6+c_6}  , 
   \end{align}
where symbols $a_i, b_i, c_i$ are intended for user 1, 2 and 3, respectively, for $i=1,2,\cdots,6$, and the power of each symbol
is $P/18$.   Then, the received signals at user~1, user~2, and use~3 take the form
\begin{align*}
  \yv_{1}  &=   \Hm_1 \Bmatrix{  a_1  \\  \vdots \\  a_6   }   +   \underbrace{\Hm_1 \Bmatrix{  b_1 \\  \vdots \\  b_6   } }_{\sv_{11}}   +    \underbrace{\Hm_1 \Bmatrix{  c_1 \\  \vdots \\  c_6   } }_{\sv_{12}} +   \zv_{1}  ,  \\
 \yv_{2}  &=      \Hm_2 \Bmatrix{  b_1 \\  \vdots \\  b_6   }    +\underbrace{ \Hm_2 \Bmatrix{  a_1  \\  \vdots \\  a_6   } }_{\sv_{21}}   +  \underbrace{ \Hm_2 \Bmatrix{  c_1 \\  \vdots \\  c_6   } }_{\sv_{22}} +   \zv_{2} ,   \\  
 \yv_{3}  &=   \Hm_3 \Bmatrix{  c_1 \\  \vdots \\  c_6   }  + \underbrace{\Hm_3 \Bmatrix{  a_1  \\  \vdots \\  a_6   } }_{\sv_{31}}  +   \underbrace{ \Hm_3 \Bmatrix{  b_1 \\  \vdots \\  b_6   }}_{\sv_{32}}    +      \zv_{3}   . 
\end{align*}
Note that the total rate of the 18 symbols \emph{overloaded} as in \eqref{eq:Ph1KMIMO} surpasses the MIMO BC capacity if each symbol carries one DoF.
One can see that, if user~1 is able to learn the variables $\sv_{11} \defeq \Hm_1 \Bmatrix{  b_1 \  \cdots \   b_6  } ^\T $, $\sv_{12} \defeq \Hm_1 \Bmatrix{  c_1 \  \cdots \   c_6  } ^\T $, $\sv_{21} \defeq \Hm_2 \Bmatrix{  a_1 \  \cdots \   a_6  } ^\T $ and $\sv_{31} \defeq \Hm_3 \Bmatrix{  a_1 \  \cdots \   a_6  } ^\T $, then user~1 can remove the interference $s_{11}$ and $s_{12}$ from $\yv_{1}$, and can use $s_{21}$ and $s_{31}$ as extra observations for decoding $a_1, \cdots, a_6$. 
Similarly, user~2 can decode $b_1, \cdots, b_6$ with the knowledge of $\sv_{11}$, $\sv_{21}$, $\sv_{22}\defeq \Hm_2 \Bmatrix{  c_1 \  \cdots \   c_6  } ^\T $, $\sv_{32}\defeq \Hm_3 \Bmatrix{  b_1 \  \cdots \   b_6  } ^\T $.  User~3 can decode $c_1, \cdots, c_6$ with the knowledge of $\sv_{12}$, $\sv_{22}$, $\sv_{31} $, $\sv_{32}$. Therefore,
in Phase~2, the transmitter will multicast  $\sv_{11},\sv_{12},\sv_{21},\sv_{22},\sv_{31}, \sv_{32}$ to all users using the multicast channel. 

\subsubsection{Phase~2 - multicast side information over the parallel multicast channel}
Phase~2 starts after the past CSI about $\Hm_1, \Hm_2, \Hm_3$ is fed back to the transmitter. 
The transmitter first \emph{regenerates} $\sv_{11}, \cdots, \sv_{32}$ based on  the past CSI, and then \emph{quantizes}
them into $\bar{\sv}_{11}, \cdots,\bar{\sv}_{32}$ by using $\rM/6$ bits for each. 
Then, the transmitter simply sends the total $\rM$ bits of the quantized values
$\bar{\sv}_{11}, \cdots,\bar{\sv}_{32}$ to all users through the multicast channel in one channel use. 
After learning $\bar{\sv}_{11}, \cdots,\bar{\sv}_{32}$, user~1, 2, 3 form their $6\times 6$ MIMO observations of the form
\begin{align*} 
\Bmatrix{ \yv_{1} -\bar{\sv}_{11}  - \bar{\sv}_{12}  \\   \bar{\sv}_{21}   \\   \bar{\sv}_{31}   }   =  \underbrace{\Bmatrix{ \Hm_1
\\ \Hm_2 \\ \Hm_3 }      \Bmatrix{ a_1 \\ \vdots \\   a_6 } }_{\text{power} \ P}
 + \underbrace{\Bmatrix {  \zv_{1}  + \tilde{\sv}_{11} + \tilde{\sv}_{12}  \\      -\tilde{\sv}_{21}  \\  -\tilde{\sv}_{31}  }  
 }_{\text{power} \ P^0} ,   
 \\  \Bmatrix{ \yv_{2} -\bar{\sv}_{21}  - \bar{\sv}_{22}  \\   \bar{\sv}_{11}   \\   \bar{\sv}_{32}   }   =  \underbrace{\Bmatrix{ \Hm_2
\\ \Hm_1 \\ \Hm_3 }      \Bmatrix{ b_1 \\ \vdots \\   b_6 } }_{\text{power} \ P}
 + \underbrace{\Bmatrix {  \zv_{2}  + \tilde{\sv}_{21} + \tilde{\sv}_{22}  \\      -\tilde{\sv}_{11}  \\  -\tilde{\sv}_{32}  }  
 }_{\text{power} \ P^0}     ,  \\  
 \Bmatrix{ \yv_{3} -\bar{\sv}_{31}  - \bar{\sv}_{32}  \\   \bar{\sv}_{12}   \\   \bar{\sv}_{22}   }   =  \underbrace{\Bmatrix{ \Hm_3
\\ \Hm_1 \\ \Hm_2 }      \Bmatrix{ c_1 \\ \vdots \\   c_6 } }_{\text{power} \ P}
 + \underbrace{\Bmatrix {  \zv_{3}  + \tilde{\sv}_{31} + \tilde{\sv}_{32}  \\      -\tilde{\sv}_{12}  \\  -\tilde{\sv}_{22}  }  
 }_{\text{power} \ P^0}  ,
\end{align*}
respectively  where $\tilde{\sv}_{i} \defeq  \sv_{i}- \bar{\sv}_{i}$ are the quantization errors. 
Since the power of $\sv_i$ is roughly $P$ and $\sv_i \in \CC^{2\times 1}$, then $\rM/6=2\log P$  bits of quantization allow for bounded power of the quantization error $\tilde{\sv}_{i}$.
Therefore,  with the help of the side information provided from the multicast channel, each user can recover its $6$ symbols from the equivalent $6\times6$ MIMO channel, achieving a sum DoF of $18$ as shown in Fig.~\ref{fig:BCSLK3N2M6}.
A simple cut-set argument reveals that even if instantaneous perfect CSIT were available at the MIMO BC transmitter, the sum DoF performance could not scale better than $18$. 
This example shows that completely outdated CSIT can be as good as instantaneous CSIT, in a sum DoF sense.

\subsection{Illustrative example $(K=3, N=1, M=3,  \dM =2/9, \ \gamma=1)$}\label{sec:SchemeExampleKMIMO2}

We now consider another example with $K=3, N=1, M=3, \rM=\frac{2}{9}\log P, \ \gamma=1$ (completely outdated CSI). Again we let $T_c=1$ for the sake of simplicity.
Different from the previous example where $\rM$ is high enough, this example has a relatively small $\rM$.

The scheme we propose for this case operates in packets of $18$ symbols in total, and each packet is transmitted over two phases, each of duration $9$ channel uses.  Specifically packet $i$ is communicated over channel uses $9i+1, 9i+2, \cdots, 9(i+1)$ of the  MIMO BC (phase~1) and channel uses $9(i+1)+1, \cdots, 9(i+2)$ of the multicast channel (phase~2), for $i=1,2,\cdots$. At the end of these two phases, each receiver can recover its $6$ symbols which yields the optimal $2$ sum DoF for the system. More precisely, in phase~1,  18 so-called order-1 symbols (each desired by only one user) are overloaded over the MIMO BC and 2 order-3 symbols (each of those symbols is desired by all the users) are generated, i.e., 2 order-3 symbols need to be transmitted to the users in order to decode those 18 order-1 symbols.
Then in phase~2 followed,  the 2 order-3 symbols are multicast over the multicast channel, which can be done in 9 channel uses since the multicast channel DoF is 
$\frac{2}{9}$.
 Next, we  describe the transmission in phase~1 and phase~2, and without loss of generality we focus on the first packet.

\subsubsection{Phase~1 -  transmit overload the MIMO BC}

The transmission in this phase is divided into two sub-phases, with durations 6 channel uses and 3 channel uses respectively.  In this specific instance, the operation in phase~1 builds on the scheme of \cite{MAT:11c} which is described below.

\begin{figure}
\centering
\includegraphics[width=10cm]{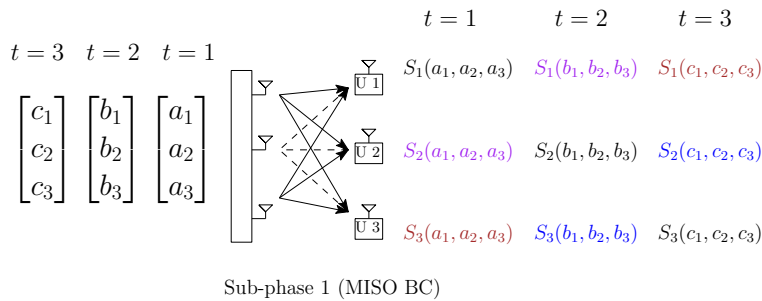}
\caption{Illustration of sub-phase~1 (of phase~1) for the proposed scheme (with $K=3, N=1, M=3, \dM =2/9, \ \gamma=1, T_c=1$), for $t=1, 2,3$ only. The transmission for $t=4, 5, 6$ is similar. }
\label{fig:K3N1d29}
\end{figure}

In sub-phase~1, the transmitter sends 18 symbols $\{a_i, b_i, c_i\}_{i=1}^{6}$ over 6 channel uses, where symbols $a_i$, $b_i$, $c_i$ are desired by user~1, user~2 and user~3 respectively (those 18 symbols are called order-1 symbols). Specifically, in the first 3 channel uses  the transmitter sends 9 symbols in the form (see Fig.~\ref{fig:K3N1d29})
\begin{align}   \label{eq:Ph1e2}
\xv [1] =  \Bmatrix{  a_1  \\  a_2 \\ a_3} , \quad  \xv [2] =  \Bmatrix{  b_1  \\  b_2 \\ b_3}, \quad \xv [3] =  \Bmatrix{  c_1  \\  c_2 \\ c_3} . 
\end{align}
Then, the received signals take the form
\begin{align*}
  y_{1} [1]  &=  \underbrace{ \hv^{\T}_1 [1] \Bmatrix{  a_1  \  a_2  \ a_3}^\T}_{S_1(a_1, a_2, a_3)}     +   z_{1}[1],  \quad   y_{1} [2]  =  \underbrace{ \hv^{\T}_1 [2] \Bmatrix{  b_1  \  b_2  \ b_3}^\T}_{S_1(b_1, b_2, b_3)}     +   z_{1}[2], \quad   y_{1} [3]  =  \underbrace{ \hv^{\T}_1 [3] \Bmatrix{  c_1  \  c_2  \ c_3}^\T}_{S_1(c_1, c_2, c_3)}     +   z_{1}[3]   ,    \\
  y_{2} [1]  &=  \underbrace{ \hv^{\T}_2 [1] \Bmatrix{  a_1  \  a_2  \ a_3}^\T}_{S_2(a_1, a_2, a_3)}     +   z_{2}[1] ,  \quad   y_{2} [2]  =  \underbrace{ \hv^{\T}_2 [2] \Bmatrix{  b_1  \  b_2  \ b_3}^\T}_{S_2(b_1, b_2, b_3)}     +   z_{2}[2], \quad   y_{2} [3]  =  \underbrace{ \hv^{\T}_2 [3] \Bmatrix{  c_1  \  c_2  \ c_3}^\T}_{S_2(c_1, c_2, c_3)}     +   z_{2}[3]  ,  \\
   y_{3} [1]  &=  \underbrace{ \hv^{\T}_3 [1] \Bmatrix{  a_1  \  a_2  \ a_3}^\T}_{S_3(a_1, a_2, a_3)}     +   z_{3} [1]  ,  \quad   y_{3} [2]  =  \underbrace{ \hv^{\T}_3 [2] \Bmatrix{  b_1  \  b_2  \ b_3}^\T}_{S_3(b_1, b_2, b_3)}     +   z_{3}[2], \quad   y_{3} [3]  =  \underbrace{ \hv^{\T}_3 [3] \Bmatrix{  c_1  \  c_2  \ c_3}^\T}_{S_3(c_1, c_2, c_3)}     +   z_{3}[3]   ,
\end{align*}
where $S_i(\bullet)$ denotes the linear function of the arguments at user~$i$.  See Fig.~\ref{fig:K3N1d29} which illustrates the first 3 channel uses.
In the next 3 channel uses,  the transmitter sends another 9 symbols $\{a_i, b_i, c_i\}_{i=4}^{6}$ in the same way as in \eqref{eq:Ph1e2}.  One can see that if user~1 is able to learn two more observations $S_2(a_1, a_2, a_3)$ and $S_3(a_1, a_2, a_3)$, then user~1 has three observations (i.e., $y_1[1]$, $S_2(a_1, a_2, a_3)$ and $S_3(a_1, a_2, a_3)$) to decode its three desired symbols $a_1, a_2, a_3$. Similarly user~2 can decode $b_1, b_2, b_3$ by learning $S_1(b_1, b_2, b_3)$ and $S_3(b_1, b_2, b_3)$, while user~3 can decode $c_1, c_2, c_3$ by learning $S_1(c_1, c_2, c_3)$ and $S_2(c_1, c_2, c_3)$.  
Therefore, in the next sub-phase the transmitter constructs these linear combinations by using its delayed CSIT and then use them to form the following $3$ order-2 symbols 
\begin{align*}
S_{AB} &\defeq S_2(a_1, a_2, a_3)  + S_1(b_1, b_2, b_3)  = \hv^{\T}_2 [1] \Bmatrix{  a_1  \  a_2  \ a_3}^\T + \hv^{\T}_1 [2] \Bmatrix{  b_1  \  b_2  \ b_3}^\T   , \\
S_{AC} &\defeq S_3(a_1, a_2, a_3)  + S_1(c_1, c_2, c_3)   = \hv^{\T}_3 [1] \Bmatrix{  a_1  \  a_2  \ a_3}^\T +  \hv^{\T}_1 [3] \Bmatrix{  c_1  \  c_2  \ c_3}^\T   ,  \\
S_{BC} &\defeq S_3(b_1, b_2, b_3)  + S_2(c_1, c_2, c_3)  =  \hv^{\T}_3 [2] \Bmatrix{  b_1  \  b_2  \ b_3}^\T +\hv^{\T}_2 [3] \Bmatrix{  c_1  \  c_2  \ c_3}^\T ,
\end{align*}
where symbol $S_{AB}$ is desired by user~1 and user~2,   $S_{AC}$ is desired by user~1 and user~3, and  $S_{BC}$ is desired by user~2 and user~3.  
To summarize in the first sub-phase of total duration $6$ channel uses, we send $9$ order-1 symbols in the  first $3$ channel uses and generate $3$ order-2 symbols ($\{S_{AB}, S_{AC}, S_{BC}\}$) to be communicated in the next sub-phase, then we send another $9$ order-1 symbols over the next $3$ channel uses and generate another $3$ order-2 symbols ($\{S_{AB}',S_{AC}', S_{BC}'\}$) again to be communicated over the next sub-phase.

\begin{figure}
\centering
\includegraphics[width=10cm]{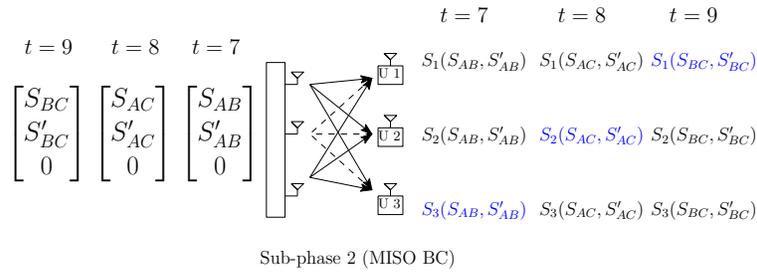}
\caption{Illustration of sub-phase~2 (of phase~1) for the proposed scheme (with $K=3, N=1, M=3, \dM =2/9, \ \gamma=1, T_c=1$).  }
\label{fig:K3N1d29subph2}
\end{figure}

In sub-phase~2, the transmitter sends the 6 order-2 symbols $\{S_{AB}, S_{AC}, S_{BC}, S_{AB}',S_{AC}', S_{BC}'\}$ over 3 channel uses in the following way (see Fig.~\ref{fig:K3N1d29subph2}):
\begin{align}   \label{eq:Ph1su2e2}
\xv [7] =  \Bmatrix{  S_{AB} \\ S_{AB}' \\ 0 } ,   \quad \xv [8] =  \Bmatrix{  S_{AC} \\ S_{AC}' \\ 0 } , \quad \xv [9] =  \Bmatrix{  S_{BC} \\ S_{BC}' \\ 0 }.
\end{align}
Here user~1 wants symbols $S_{AB}, S_{AC},S_{AB}',S_{AC}'$, user~2 wants symbols $S_{AB}, S_{BC},S_{AB}',S_{BC}'$, and user~3 wants symbols $S_{AC}, S_{BC},S_{AC}',S_{BC}'$. Then, the received signals take the form
\begin{align*}
  y_{1} [7]  \!&=\!  \underbrace{ \hv^{\T}_1 [7] \!\Bmatrix{  S_{AB}  \  S_{AB}'  \ 0}^\T}_{S_1(S_{AB}, S_{AB}')}     \!+\!   z_{1}[7],  \quad   y_{1} [8]  \!=\!  \underbrace{ \hv^{\T}_1 [8]\! \Bmatrix{  S_{AC}  \  S_{AC}'  \ 0 }^\T}_{S_1(S_{AC}, S_{AC}')}     \!+\!   z_{1}[8], \quad   y_{1} [9]  \!=\!  \underbrace{ \hv^{\T}_1 [9] \!\Bmatrix{  S_{BC}  \  S_{BC}'  \ 0}^\T}_{S_1(S_{BC}, S_{BC}')}     \!+\!   z_{1}[9]    ,   \\
  y_{2} [7]  \!&=\!  \underbrace{ \hv^{\T}_2 [7]\! \Bmatrix{  S_{AB}  \  S_{AB}'  \ 0 }^\T}_{S_2(S_{AB}, S_{AB}')}     +   z_{2}[7] ,  \quad   y_{2} [8]  \!=\!  \underbrace{ \hv^{\T}_2 [8]\! \Bmatrix{  S_{AC}  \  S_{AC}'  \ 0}^\T}_{S_2(S_{AC}, S_{AC}')}    \! + \!  z_{2}[8], \quad   y_{2} [9]  \!=\!  \underbrace{ \hv^{\T}_2 [9]\! \Bmatrix{  S_{BC}  \  S_{BC}'  \ 0}^\T}_{S_2(S_{BC}, S_{BC}')}     \!+\!   z_{2}[9]   , \\
   y_{3} [7]  \!&=\!  \underbrace{ \hv^{\T}_3 [7]\! \Bmatrix{  S_{AB}  \  S_{AB}'  \ 0}^\T}_{S_3(S_{AB}, S_{AB}')}     +   z_{3} [7]  ,  \quad   y_{3} [8]  \!=\!  \underbrace{ \hv^{\T}_3 [8] \!\Bmatrix{  S_{AC}  \  S_{AC}'  \ 0}^\T}_{S_3(S_{AC}, S_{AC}')}     \!+\!   z_{3}[8], \quad   y_{3} [9]  \!=\!  \underbrace{ \hv^{\T}_3 [9]\! \Bmatrix{  S_{BC}  \  S_{BC}'  \   0 }^\T}_{S_3(S_{BC}, S_{BC}')}     \!+\!   z_{3}[9]  .
\end{align*}        
Note that if each user has the knowledge of the following two order-3 symbols
\begin{align*}
S_{ABC} &\defeq         \beta_{1,1} S_3(S_{AB}, S_{AB}')  + \beta_{1,2}   S_2(S_{AC}, S_{AC}') +  \beta_{1,3}    S_1(S_{BC}, S_{BC}')  , \\
S_{ABC}' &\defeq     \beta_{2,1} S_3(S_{AB}, S_{AB}')  + \beta_{2,2}   S_2(S_{AC}, S_{AC}') +  \beta_{3,3}    S_1(S_{BC}, S_{BC}')  ,
\end{align*}
where $\beta_{i,j}$, $i = 1,2$, $j=1,2,3$,   are constants  that we assume have been shared between all the nodes ahead of time,  then  each user can decode its desired order-2 symbols. 
Therefore, in this sub-phase~2, we send $6$ order-2 symbols and generate $2$ order-3 symbols. As we discuss next, these $2$ order-3 symbols are sent through the multicast channel in the following phase.

\begin{figure}
\centering
\includegraphics[width=4cm]{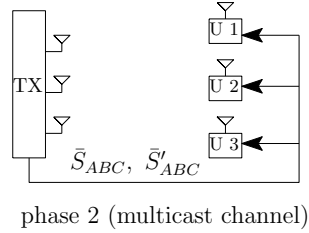}
\caption{Illustration of phase~2 for the proposed scheme (with $K=3, N=1, M=3, \dM =2/9, \ \gamma=1, T_c=1$).  }
\label{fig:K3N1d29ph2}
\end{figure}

\subsubsection{Phase~2 - multicast side information over the parallel multicast channel}
Phase~2 operates in 9 channel uses over the multicast channel.
The transmitter first \emph{regenerates} $S_{ABC}, S_{ABC}'$ based on  the past CSI, and then \emph{quantizes} them into $\bar{S}_{ABC},  \bar{S}_{ABC}'$ by using $2\log P$ bits in total, such that the quantization error is under the noise level. 
Then, the transmitter simply sends the total $2\log P$ bits of the quantized values
$\bar{S}_{ABC},  \bar{S}_{ABC}'$ to all users through the multicast channel in $9$ channel uses, since the capacity of the multicast channel is $\rM=2/9\log P$ bits/channel use. See Fig.~\ref{fig:K3N1d29ph2}.  
After learning $\bar{S}_{ABC},  \bar{S}_{ABC}'$, user~1, 2, 3 can decode their own order-2 symbols and then decode their own order-1 symbols as mentioned in phase~1.

Therefore,  with the help of the side information provided over the multicast channel, each user can receive $6$ symbols in every 9 channel uses, achieving a sum DoF of $\frac{3\times 6}{9}=2$. 
It turns out that this is the optimal sum DoF that we can get in this case (cf. Proposition~\ref{pro:KMIMOsum})

Note that the two presented examples are based on  a two-phase overload-multicast strategy. 
However, for the previous example with large enough $\dM$  ($\dM \geq K(K-1)N\gamma$),  the rate of the overall symbols transmit overloaded over the MIMO BC is scaled with  $K^2N\log P$ bits per channel use  (so called fully transmit overload); while for this example with small  $\dM$  ($\dM < K(K-1)N\gamma$),   the rate of the transmit overload symbols  is scaled less than  $K^2N\log P$ bits per channel use  (so called partially transmit overload). Specifically, with partially transmit overload, the rate of the information needed to multicast over the multicast channel is  reduced (compared with that corresponding to fully transmit overload),  consequently matching the capacity of the multicast channel.

\section{Conclusion}   

This work characterizes the optimal DoF region of the two-user MIMO BC with a multicast channel as a function of two parameters: the multicast channel capacity  and the CSIT timeliness  for the MIMO BC. The result reveals that completely outdated CSIT can achieve the same sum DoF performance as with instantaneous CSIT  if the multicast channel capacity is above a certain threshold. More precisely, there is an inherent tradeoff between the CSIT timeliness and the multicast channel capacity: with almost timely CSIT a small multicast channel capacity is enough to achieve the instantaneous CSIT performance; with completely outdated CSIT a large multicast channel capacity is required to compensate for the sum DoF loss due to the CSIT staleness.    

The optimal sum DoF is achieved by a two-phase overload-multicast strategy. The main idea of this strategy is to send information over the MIMO BC at a rate above its capacity and  use the multicast channel to send additional information to enable reliable decoding.  
The same strategy extends to the $K$-users MIMO BC with a  parallel multicast channel, and is shown again to achieve instantaneous CSIT performance, in the sum DoF sense, with completely outdated CSIT provided that the multicast channel capacity is large enough. When $K$ is large, the sum DoF gain of the proposed joint coding strategy over seperate coding over the two parallel channels is proportional to the total number of receive antennas. 

The setup we consider here arises in heterogeneous networks where transmitters and receivers are connected over multiple networks. Our work reveals that joint coding over such networks can provide significant gain in capacity. This is in sharp contrast to the well-known results on traditional parallel BCs where parallel channels are formed by different time/frequency realizations of the same physical channel. While using individually optimized codes for each channel is optimal in this case, our result reveals that for heterogeneous parallel channels joint coding may be needed. Our future work will focus on exploring optimal communication over other heterogeneous networks.

\appendices

\section{Achievability details for the $K$-user case} \label{sec:generalsch}

In this section we provide the achievability details for the $K$-user $K\times M \times N$ $(M\geq KN)$ MIMO BC with a multicast channel. 
The illustrative schemes in Section~\ref{sec:SchemeExampleKMIMO} and Section~\ref{sec:SchemeExampleKMIMO2} were designed for a particular choice of $\gamma, K, \dM, M, N$. Here we describe the general scheme for arbitrary values of these parameters. Specifically we show that the following DoF points are achievable:
\begin{align}
Q_o \defeq \Bigl(\dM   &=   K(K-1)N\gamma,  \quad \dsum = KN + \dM  \Bigr)   ,   \label{eq:defQo}\\ 
Q_L \defeq \Bigl(\dM  &=\frac{ (K-L)N\gamma}{ (L+1) \sum_{k=1}^{L}  \frac{1}{ k}},  \quad \dsum =\frac{ KN\gamma}{ \sum_{k=1}^{L}  \frac{1}{ k}}    +  (1-\gamma) KN \Bigr),  \quad   \text{for}  \quad  L=1,2,\cdots,K    . \label{eq:defQL}
\end{align}
As we will show later on, time sharing between these points achieves the whole region stated in  Proposition~\ref{pro:KMIMOBC}.

\subsection{Achieving $Q_o$}  
In this case $\dM  =   K(K-1)N\gamma$ is large enough and we show that the sum DoF $\dsum = KN + \dM = K^2N\gamma + KN(1-\gamma) $ is achievable.  The scheme which achieves this point is the extension of the example in Section~\ref{sec:SchemeExampleKMIMO}. We summarize the following basic principles for this scheme. 
\begin{itemize}
\item When instantaneous CSIT is available over the MIMO BC (over a $1- \gamma$ fraction of the block), $KN$ fresh symbols are sent with spatial zero-forcing precoding, allowing each user to decode its corresponding $N$ symbols in one channel use.  

\item When instantaneous CSIT is not available (over a $\gamma$ fraction of the block),  the transmitter overloads the MIMO BC, i.e., it transmits $K^2N$ fresh symbols in one channel use by using the signaling technique suggested in \eqref{eq:Ph1KMIMO}. 

\item In order to decode these $K^2N$ symbols, the transmitter needs to multicast extra side information of $K(K-1)N\log P$ bits to all users, and does so over the multicast channel (see the example of Section~\ref{sec:SchemeExampleKMIMO}).
\end{itemize}

Note that the amount of side information generated in one block given by $$K(K-1)N\gamma\log P$$ matches exactly the total multicasting capacity in the next block, i.e., $K(K-1)N\gamma = \dM$. As a result the following sum DoF is achievable
\begin{align*}
\dsum= \underbrace{K^2N \gamma }_{\text{overloading}}  +    \underbrace{KN(1-\gamma ) }_{\text{ZF}} .
\end{align*} 
Note that during overloading we transmit $K^2N$ symbols, hence achieve $K^2N$ DoF once  the extra side information is obtained by the receivers and the interference is resolved. Since side information multicasting does not provide any fresh information it does not contribute to the DoF computation given above.

\subsection{Achieving  $Q_L$}
We next show that,   given $\dM=\frac{N\gamma (K-L) }{(L+1)\sum_{k=1}^{L} \frac{1}{k}  }$, the sum DoF $\dsum = \frac{ KN\gamma}{ \sum_{k=1}^{L}  \frac{1}{ k}}    +  (1-\gamma) KN $ is achievable, for $L=1,2,\cdots,K$.  The scheme is the extension of the example in Section~\ref{sec:SchemeExampleKMIMO2}. We summarize the following basic principles for this scheme. 

\begin{table}
\caption{Phase~1 summary of the scheme for achieving $Q_L$. }
\begin{center}
{\renewcommand{\arraystretch}{1.5}
\begin{tabular}{|c|c|c|c|}
  \hline
   sub-phase $j$     & order-$j$ symbols sent &  order-$(j+1)$ symbols generated   & used time (channel uses) \\
   \hline
   $j=1$    &   $ K N\gamma   {K \choose 1} / {K -1 \choose 0 } $   &  $ N\gamma  {K \choose 2 } / {K -1 \choose 0 } $  &  $ \gamma {K \choose 1} / {K -1 \choose 0} $   \\
   \hline
  $j=2$    &   $ (K-1) N\gamma   {K \choose 2} / {K -1 \choose 1} $   &  $ 2 N\gamma {K \choose 3} / {K -1 \choose 1 } $  &  $ \gamma {K \choose 2} / {K -1 \choose 1} $   \\
   \hline
  $\vdots$   &   $\vdots$   &  $\vdots$  &  $\vdots$  \\
   \hline
  $j=L$   &   $ (K+1-L)N\gamma   {K \choose L} / {K -1 \choose L - 1} $   &  $ L N\gamma  {K \choose L + 1 } / {K -1 \choose L - 1} $  &  $ \gamma {K \choose L} / {K -1 \choose L - 1} $   \\
   \hline
\end{tabular}
}
\end{center}
\label{tab:ph1summary}
\end{table}

\begin{itemize}
\item When instantaneous CSIT is available over the MIMO BC (over a $1- \gamma$ fraction of the block), $KN$ fresh symbols are sent with spatial zero-forcing precoding allowing each user to decode its corresponding $N$ symbols in one channel use.  

\item When instantaneous CSIT is not available (over a $\gamma$ fraction of the block),  the transmitter follows the two-phase strategy illustrated in Section~\ref{sec:SchemeExampleKMIMO2}. Specifically, as shown in Table~\ref{tab:ph1summary}, phase~1 consists of $L$ sub-phases over the MIMO BC. In sub-phase~$j$, for $j=1,2,\cdots,L$, the transmitter sends order-$j$ symbols.  Following  \cite{MAT:11c},  sub-phase~$j$ has duration  \[  T_u (j) \defeq \frac{\gamma {K \choose j} }{ {K -1 \choose j - 1} }    \] channel uses and the transmitter sends $\phi_s (j)$ number of order-$j$ symbols where \[   \phi_s (j) \defeq  \frac{ (K+1-j)N\gamma   {K \choose j} }{  {K -1 \choose j - 1} } \ ,\] and generates $\phi_g (j)$ number of order-$(j+1)$ symbols to be sent in the next sub-phase (in order to decode those order-$j$ symbols) where \[ \phi_g (j) \defeq \frac{ j N\gamma  {K \choose j + 1 } }{ {K -1 \choose j - 1} } \ .\] 
Note that the number of the order-$(j+1)$ symbols generated in sub-phase~$j$ matches the number of order-$(j+1)$ symbols sent in sub-phase~$(j+1)$, i.e.,  $\phi_g (j) =  \phi_s (j+1)$  for $j=1,2,\cdots,L-1$.

\item At the end of sub-phase~$L$,  $\phi_g (L)$ number of order-$(L+1)$ symbols are generated. The communication of this symbols is delegated to the multicast channel. More precisely, in phase 2 of the scheme the quantized versions of the order-$(L+1)$ symbols are multicast over the multicast channel finally allowing to decode the desired order-1 symbols at all the receivers. 
\end{itemize}
 
Note that the rate of the order-$(L+1)$ symbols to be multicast matches the rate of the multicast channel, i.e., \[ \frac{\phi_g (L)}{ \frac{1}{\gamma} \sum_{j=1}^{L} T_u (j)} =  \frac{N\gamma (K-L) }{(L+1)\sum_{k=1}^{L} \frac{1}{k}  } = \dM  .  \]  
Note that $\frac{1}{\gamma} \sum_{j=1}^{L} T_u (j)$ is the total length of each communication block as in Figure~\ref{fig:BlocksCoop}. (Over the MIMO BC, this block is shared between spatial zero forcing and phase 1 of the above scheme.) As a result the following sum DoF is achievable
\begin{align*}
\dsum= \underbrace{\frac{  \phi_s (1)  }{ \frac{1}{\gamma} \sum_{j=1}^{L} T_u (j)}  }_{\text{overloading}}  +    \underbrace{KN(1-\gamma ) }_{\text{ZF}}  =\frac{ KN\gamma}{ \sum_{k=1}^{L}  \frac{1}{ k}}    +  (1-\gamma) KN .
\end{align*}

\subsection{Achieving  intermediate points}

\begin{figure}
\centering
\includegraphics[width=10cm]{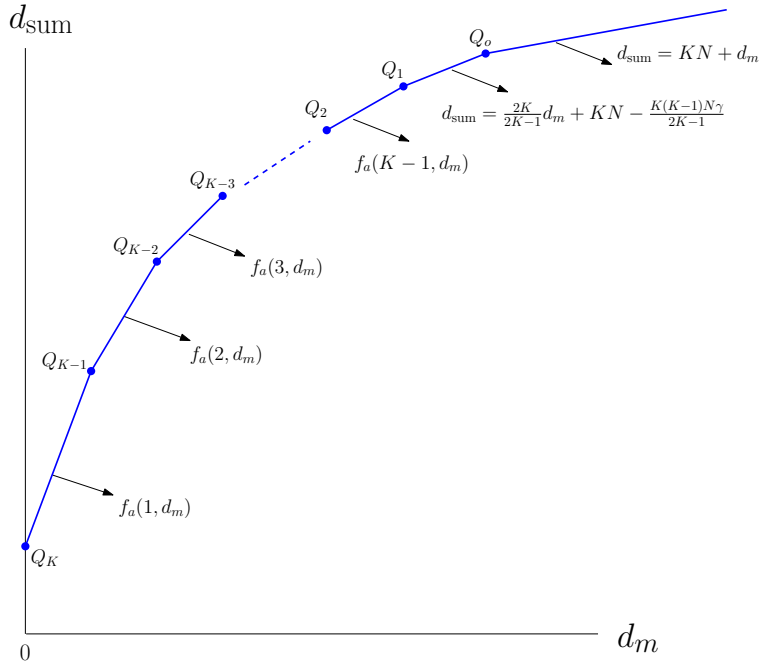}
\caption{DoF inner bounds illustration for the $K$-user MIMO BC with a multicast channel. }
\label{fig:SumDoFKinnerBound}
\end{figure}

First we show that time sharing between two strategies achieving DoF points
$Q^{\star} = (\dM^{\star} , \dsum^{\star})$ and $Q^{'} = (\dM^{'}, \dsum^{'})$ respectively, for $ \dM^{\star}<\dM^{'}$,  gives the following sum DoF point
\begin{align}    \label{eq:linege}
\dsum  =   \dsum^{\star}   +  \frac{(\dM-\dM^{\star}  ) ( \dsum^{'} - \dsum^{\star}  ) }{\dM^{'} - \dM^{\star}  }  ,      \    \text{for}   \ \    \dM^{\star}   \leq  \dM   \leq  \dM^{'} .
 \end{align}
To achieve the above point, the time fractions allocated to the first  strategy (achieving $Q^{\star}$) and the second strategy (achieving $Q^{'}$) are chosen as
\[  1-\Delta_t    \quad  \text{and}    \quad    \Delta_t \defeq \frac{\dM-\dM^{\star}   }{\dM^{'} - \dM^{\star}  }  , \] 
respectively, which allows to achieve the DoF in \eqref{eq:linege}  with  the following DoF for the multicast channel 
\[    \Delta_t \dM^{'}  + (1-\Delta_t) \dM^{\star}   =\dM^{\star}   + \Delta_t (\dM^{'} - \dM^{\star}  )  = \dM^{\star}   +   \frac{\dM-\dM^{\star}   }{\dM^{'} - \dM^{\star}  }  (\dM^{'} - \dM^{\star}  )  = \dM. \] 
Following the same argument, time sharing between two strategies achieving DoF points
$Q_L$ and $Q_{L-1}$ (cf. \eqref{eq:defQL}) gives the following sum DoF performance 
\begin{align}    \label{eq:lineL}
    \dsum  =  \frac{ K(L+1)L \dM  +  KN\gamma (K\!+\!1) L }{ (L\!+\!1)(K\!+\!1\!-\!L) \!+\! (K\!+\!1)L \sum_{k=1}^{L-1} \frac{1}{k}  }  + KN(1\!-\!\gamma),     \    \text{for}   \ \    \frac{ (K-L)N\gamma}{ (L\!+\!1) \sum_{k=1}^{L}  \frac{1}{ k}}   \!\leq\! \dM  \!\leq \! \frac{ (K\!-\!L\!+\!1)N\gamma}{ L \sum_{k=1}^{L-1}  \frac{1}{ k}}    ,
 \end{align}
 for $L= 2, 3, \cdots, K$.  Note that the expression in \eqref{eq:lineL} can be equivalently written as
\begin{align}  
   \dsum =  f_a(K+1-L, \dM) ,     \    \text{for}   \ \     f_b ( K+1-L )   \leq \dM  \leq  f_b ( K+2- L )   ,  
 \end{align}
which matches the third line in Proposition~\ref{pro:KMIMOBC}   (see \eqref{eq:BCKfa}, \eqref{eq:BCKfb}, \eqref{eq:JCkmimobc}, and see Fig.~\ref{fig:SumDoFKinnerBound}).   

Similarly time sharing between two strategies achieving DoF points
$Q_1$ and $Q_o$ gives the following sum DoF performance (cf. \eqref{eq:JCkmimobc})
\begin{align}    \label{eq:line1}
\dsum  =  \frac{2K}{2K-1} \dM +  KN -  \frac{K(K-1)N\gamma}{2K-1},     \    \text{for}   \ \    \frac{ (K-1)N\gamma}{ 2}   \leq  \dM   \leq    K(K-1)N\gamma  ,
 \end{align}
which matches the second line in Proposition~\ref{pro:KMIMOBC}.

Finally  the sum DoF performance 
\begin{align}    \label{eq:line0}
\dsum = KN + \dM    \quad    \text{for}   \quad   \dM  \geq K(K-1)N\gamma ,
 \end{align}
is achievable by applying the strategy that achieves the sum DoF point $Q_o$.  Note that using this strategy, a sum DoF of $KN + K(K-1)N\gamma$ is achievable  when $\dM = K(K-1)N\gamma$. When $\dM$ is larger than the threshold $K(K-1)N\gamma$, then the remaining DoF $\dM - K(K-1)N\gamma$ on the multicast channel can be used for transmitting an independent message over the multicast channel, which allows us to achieve the sum DoF in \eqref{eq:line0}, i.e., 
\begin{align*}   
\dsum = \underbrace{KN + K(K-1)N\gamma}_{\text{strategy for}  \ Q_o} +  \underbrace{ \dM - K(K-1)N\gamma}_{\text{independent transmission}}  = KN + \dM     \quad    \text{for}   \quad   \dM  \geq K(K-1)N\gamma. 
 \end{align*}
This completes the proof of  Proposition~\ref{pro:KMIMOBC}.

\section{Converse for the $K$-user case   \label{sec:OuterbKMIMOBC}}

In this section we provide the converse proof for the $K$-user MIMO BC with a multicast channel (cf. Proposition~\ref{pro:KMIMOsum}).  
The proof is based on Fano's inequality, basic entropy inequalities, genie-aided  techniques, as well as the symmetric entropy technique. 
For the $K$-user case, it suffices to prove the following lemma.

\vspace{5pt}
\begin{lemma}  \label{lm:KMIMODoFregion}
For the $K$-user $K\times M \times N$ MIMO BC with a limited-rate  multicast  channel, the  DoF region is upper bounded as
\begin{align}
\sum_{k=1}^{K}  \frac{d_{\pi(k) }}{ k }   &\leq \dM + N\gamma + N(1-\gamma)  \sum_{k=1}^{K}  \frac{1}{ k} ,  \quad \forall  \  \pi   ,  \label{eq:KMboundsum}  \\
\sum_{k\in \{1, 2, \cdots, L\}}  \!\!\!\! d_{\pi(k) }    +  \frac{L}{K} \!\!\! \sum_{j \in \{1,2,\cdots,K \} \backslash \{1,2,\cdots, L \}}    \!\!\!\!\!\!\!\!\!\! d_{ \pi(j)  }   &\leq      \dM +NL  + \frac{NL(K\!-\!L)(1\!-\! \gamma)}{K}   ,  \quad \forall  \  \pi, \  \text{for} \  L=1, \cdots, K   ,  \label{eq:boundK1}
\end{align}
where $\pi$ denotes a permutation of the set $\{1,2,\cdots, K\}$,  and $\pi(k)$ denotes the $k$th element of the permuted set.  
\end{lemma}

Note that the first bound of Proposition~\ref{pro:KMIMOsum} (cf. \eqref{eq:BCKfa}, \eqref{eq:OBkmimobc}) \[d_{\text{sum}}   \leq   \frac{ K \dM+ KN\gamma}{ \sum_{k=1}^{K}  \frac{1}{ k} } + KN(1-\gamma) ,  \] follows from \eqref{eq:KMboundsum}   (by summing $K$ different bounds as in \eqref{eq:KMboundsum}), while  the  second bound (cf. \eqref{eq:BCKfp}, \eqref{eq:OBkmimobc}) 
\[d_{\text{sum}}   \leq  \frac{ K^2   \dM }{KL+L(K-L)}  +\frac{ K^2 NL + KNL(K-L)(1-\gamma) }{KL+L(K-L)} ,  \] 
follows from \eqref{eq:boundK1}  (by summing $K$ different bounds as in \eqref{eq:boundK1}).

Before showing the proof details, we provide one lemma to be used. Note that this lemma is a generalized result of  Lemma~\ref{lm:DiffEntroMIMOBC} based on the entropy symmetry. 
\vspace{5pt}
\begin{lemma}  \label{lm:DiffEntroKMIMO} 
 $   h(\yv_1^n,\ldots,\yv_{K}^n
  \cond W_1,\ldots,W_{L}, \Omega^n) - \frac{K}{L}
  h(\yv_1^n,\ldots,\yv_L^n \cond W_1,\ldots,W_{L}, \Omega^n) \leq   n(1-\gamma)(K-L) \bigl(N\log P+o(\log P)\bigr) $ for $L=1,2,\cdots,K-1$. 
\end{lemma}
\vspace{5pt}
\begin{proof}
We again let $\TN \defeq \left\{ t\in[1,n]: \, \text{current channel state
is not known at time $t$} \right\}$, with $|\TN|=n\gamma$.  Let $\Uc \defeq  \{W_1,\ldots,W_{K}, \Omega^n\}$. Then, we have
\begin{align} 
&   h(\yv_1^n,\ldots,\yv_{K}^n \cond \Uc ) - \frac{K}{L}  h(\yv_1^n,\ldots,\yv_L^n \cond \Uc )  \nonumber\\
&= \sum_{t=1}^n \Bigl(h(\yv_{1}[t], \ldots,\yv_{K}[t] \cond \yv^{t-1}_{1}, \ldots, \yv^{t-1}_{K}, \Uc)   -   \frac{K}{L}  h(\yv_{1}[t],\ldots, \yv_{L}[t] \cond\yv^{t-1}_{1}, \ldots, \yv^{t-1}_{L},\Uc ) \Bigr)     \label{eq:b0928}
 \\
&\leq \sum_{t=1}^n \Bigl(h(\yv_{1}[t], \ldots,\yv_{K}[t] \cond \yv^{t-1}_{1}, \ldots, \yv^{t-1}_{L}, \Uc)   -   \frac{K}{L}  h(\yv_{1}[t],\ldots, \yv_{L}[t] \cond\yv^{t-1}_{1}, \ldots, \yv^{t-1}_{L},\Uc ) \Bigr)     \label{eq:b9728}
 \\
&=  \sum_{t \in \TN } \Bigl(h(\yv_{1}[t], \ldots,\yv_{K}[t] \cond \yv^{t-1}_{1}, \ldots, \yv^{t-1}_{L}, \Uc)   -   \frac{K}{L}  h(\yv_{1}[t],\ldots, \yv_{L}[t] \cond\yv^{t-1}_{1}, \ldots, \yv^{t-1}_{L},\Uc ) \Bigr)    \nonumber\\  & \quad  + \sum_{t\not\in \TN } \Bigl(h(\yv_{1}[t], \ldots,\yv_{K}[t] \cond \yv^{t-1}_{1}, \ldots, \yv^{t-1}_{L}, \Uc)   -   \frac{K}{L}  h(\yv_{1}[t],\ldots, \yv_{L}[t] \cond\yv^{t-1}_{1}, \ldots, \yv^{t-1}_{L},\Uc ) \Bigr)    \nonumber \\
 & \leq    \sum_{t\not\in \TN } \Bigl(h(\yv_{1}[t], \ldots,\yv_{K}[t] \cond \yv^{t-1}_{1}, \ldots, \yv^{t-1}_{L}, \Uc)   -   \frac{K}{L}  h(\yv_{1}[t],\ldots, \yv_{L}[t] \cond\yv^{t-1}_{1}, \ldots, \yv^{t-1}_{L},\Uc ) \Bigr)  \label{eq:bsymen}   \\
 & =    \sum_{t\not\in \TN } \Bigl(  h( \yv_{L+1}[t], \ldots,\yv_{K}[t] \cond   \yv_{1}[t], \ldots, \yv_{L}[t], \yv^{t-1}_{1}, \ldots, \yv^{t-1}_{L}, \Uc)   \nonumber\\  & \quad\quad \quad \quad  -  \frac{K-L}{L}   h(\yv_{1}[t],\ldots, \yv_{L}[t] \cond\yv^{t-1}_{1}, \ldots, \yv^{t-1}_{L},\Uc ) \Bigr)  \label{eq:chain}   \\
 &\leq    (n - |\TN|) (K-L)\bigl(N\log P+o(\log P) \bigr)    ,   \label{eq:KMMcondentro432}   
\end{align}
where \eqref{eq:b0928}  follows from the basic chain rule,  \eqref{eq:b9728} uses  the fact that conditioning reduces differential entropies;   \eqref{eq:bsymen}  is due to the symmetry of the output whenever the channel input is independent of the current channel state,  i.e.,  $h(\yv_{1}[t], \ldots,\yv_{K}[t] \cond \yv^{t-1}_{1}, \ldots, \yv^{t-1}_{L}, \Uc)   \leq   \frac{K}{L}  h(\yv_{1}[t],\ldots, \yv_{L}[t] \cond\yv^{t-1}_{1}, \ldots, \yv^{t-1}_{L},\Uc ) $ when $t\in \TN$  (cf.~\cite[Lemma~3]{YKPS:14}), 
\eqref{eq:chain} uses the basic chain rule, and the last inequality holds since
$h( \yv_{L+1}[t], \ldots,\yv_{K}[t] \cond   \yv_{1}[t], \ldots, \yv_{L}[t], \yv^{t-1}_{1}, \ldots, \yv^{t-1}_{L}, \Uc)   \le N(K-L)\log P+o(\log P)$ and 
$ h(\yv_{1}[t],\ldots, \yv_{L}[t] \cond\yv^{t-1}_{1}, \ldots, \yv^{t-1}_{L},\Uc ) \ge h(\zv_{1}[t],\ldots, \zv_{L}[t]) = o(\log P)$. 
Finally, by subsisting  $|\TN|$ with $n\gamma$, we complete the proof. 
\end{proof}
\vspace{5pt}

In what follows, we provide the proofs for bounds \eqref{eq:KMboundsum} and \eqref{eq:boundK1}.

\subsection{Proof of bound \eqref{eq:KMboundsum}}

We first prove the bound in \eqref{eq:KMboundsum}. Without loss of generality, we focus on the case with $\pi : = \{1,2,\cdots,K\}$, while the other cases follow easily due to the symmetry.
By providing the observations and messages of users $1, 2, \ldots,k-1$ to user
$k$, we derive the following genie-aided upper bounds on the achievable
rates 
\begin{align}
  n R_1 &\le I(W_1; \yv^{n}_1, y^{n}_0 | \Omega^{n})  +  n \epsilon_n ,  \\
  n R_2 &\le I(W_2; \yv^{n}_1, \yv^{n}_2, y^{n}_0 | W_1, \Omega^{n})  +  n \epsilon_n ,   \\
  &\ \vdots \nonumber \\
  n R_K &\le I(W_K; \yv^{n}_1, \yv^{n}_2,\ldots, \yv^{n}_K,y^{n}_0 | W_1, \ldots,W_{K-1},\Omega^{n})  +  n \epsilon_n    , 
\end{align}
by applying Fano's inequality and some basic chain rules, and using the independence between the messages.
Then, we have
\begin{align}
  & n R_k -n \epsilon_n  \nonumber\\
  & \le I(W_k; \yv^{n}_1, \yv^{n}_2,\ldots, \yv^{n}_k,y^{n}_0 | W_1, \ldots,W_{k-1},\Omega^{n})      \nonumber\\
  & =  I(W_k; \yv^{n}_1, \yv^{n}_2,\ldots, \yv^{n}_k | W_1, \ldots,W_{k-1},\Omega^{n})  + I(W_k; y^{n}_0 | \yv^{n}_1,\ldots, \yv^{n}_k, W_1, \ldots,W_{k-1},\Omega^{n})    \nonumber\\
  & =  h(\yv^{n}_1, \yv^{n}_2,\ldots, \yv^{n}_k | W_1, \ldots,W_{k-1},\Omega^{n})  -h(\yv^{n}_1, \yv^{n}_2,\ldots, \yv^{n}_k | W_1, \ldots,W_{k},\Omega^{n})   \nonumber\\ & \quad +   H(y^{n}_0 | \yv^{n}_1,\ldots, \yv^{n}_k,W_1, \ldots,W_{k-1},\Omega^{n}) -H(y^{n}_0 | \yv^{n}_1,\ldots, \yv^{n}_k,W_1, \ldots,W_{k},\Omega^{n})   ,     \label{eq:Rkbound}
\end{align}
for $k=1,2, \cdots,K$, where $\{W_1, \ldots,W_{k-1}\}$ denotes an empty set when $k=1$.
From \eqref{eq:Rkbound},  we consequently  have 
\begin{align}
  &\sum_{k=1}^K \frac{n}{k} (R_k-\epsilon_n)     \nonumber \\
  &\le  \sum_{k=1}^{K-1} \underbrace{ \left(  \frac{1}{k+1} h(\yv_1^n,\ldots,\yv_{k+1}^n
  \cond W_1,\ldots,W_{k}, \Omega^n) - \frac{1}{k}
  h(\yv_1^n,\ldots,\yv_k^n \cond W_1,\ldots,W_{k}, \Omega^n) \right)}_{\leq\frac{n(1-\gamma)}{k+1}  \bigl(N\log P+o(\log P)\bigr)}
  \nonumber \\
  &\qquad + h(\yv_1^n \cond \Omega^n) -
  \frac{1}{K} h(\yv_1^n,\ldots,\yv_K^n \cond W_1,\ldots,W_{K},\Omega^n)  \nonumber\\
 &  \qquad + \sum_{k=1}^{K-1} \left( \frac{1}{k+1}  \underbrace{ H( y^{n}_0
  \cond \yv^{n}_1,\ldots, \yv^{n}_{k+1},W_1,\ldots,W_{k}, \Omega^n)}_{\leq H( y^{n}_0
  \cond \yv^{n}_1,\ldots, \yv^{n}_{k},W_1,\ldots,W_{k}, \Omega^n)  } - \frac{1}{k}
  H(y^{n}_0 \cond \yv^{n}_1,\ldots, \yv^{n}_k, W_1,\ldots,W_{k}, \Omega^n) \right)
  \nonumber \\
  &\qquad + H(y^{n}_0 \cond \yv^{n}_1,\Omega^n) -
  \frac{1}{K} H(y^{n}_0 \cond \yv^{n}_1,\ldots, \yv^{n}_K, W_1,\ldots,W_{K},\Omega^n) \nonumber \\
&\le  \sum_{k=1}^{K-1} \left( \frac{n(1-\gamma)}{k+1}  \bigl(N\log P+o(\log P)\bigr)\right) +\underbrace{ h(\yv_1^n \cond \Omega^n)}_{\leq n N \log P +n\cdot o(\log P)} -
  \frac{1}{K} \underbrace{ h(\yv_1^n,\ldots,\yv_K^n \cond W_1,\ldots,W_{K},\Omega^n)}_{\geq  n\cdot o(\log P) }  \nonumber\\
  &  \qquad + \sum_{k=1}^{K-1}  \left( \underbrace{\Bigl(\frac{1}{k+1} - \frac{1}{k} \Bigr) }_{ < 0}  \underbrace{H( y^{n}_0
  \cond \yv^{n}_1,\ldots, \yv^{n}_{k},W_1,\ldots,W_{k}, \Omega^n)}_{ \geq 0} \right)
  \nonumber \\
  &\qquad + \underbrace{H(y^{n}_0 \cond \yv^{n}_1,\Omega^n)}_{\leq n\rM} -
  \frac{1}{K} \underbrace{H(y^{n}_0 \cond \yv^{n}_1,\ldots, \yv^{n}_K, W_1,\ldots,W_{K},\Omega^n)}_{ \geq 0}   \label{eq:up314}\\
&\le  \sum_{k=1}^{K-1} \left( \frac{nN(1-\gamma)}{k+1}\log P  \right) +nN\log P +n\cdot o(\log P)  + n\rM    ,  \label{eq:up4243}
\end{align}
where \eqref{eq:up314} is from Lemma~\ref{lm:DiffEntroKMIMO}  and the fact that conditioning reduces entropy, and the last inequality follows from  the non-negativity of the entropy, and that $H(y^{n}_0 \cond \yv^{n}_1,\Omega^n) \leq n\rM$ and $h(\yv_1^n,\ldots,\yv_K^n \cond W_1,\ldots,W_{K},\Omega^n) \geq h(\yv_1^n,\ldots,\yv_K^n \cond \xv^n, W_1,\ldots,W_{K},\Omega^n) = h(\zv_1^n,\ldots,\zv_K^n ) =  n\cdot o(\log P)$.  
Hence, dividing \eqref{eq:up4243} by $n \log P$ and let $P\to\infty$, \eqref{eq:KMboundsum}
follows immediately.

\subsection{Proof of bound \eqref{eq:boundK1}}

Now we prove the bound in \eqref{eq:boundK1}, and again without loss of generality we focus on the case with $\pi : = \{1,2,\cdots,K\}$.
We  at first consider the case with $L<K$, and then consider the case with $L=K$ later on.
As the first step,  we enhance the original BC by allowing cooperation between the first $L$ users (consequently each of users $1, 2, \ldots, L$ observes channel outputs $\yv^{n}_1,\yv^{n}_2, \cdots, \yv^{n}_{L}, y^{n}_0$), for $1\leq L \leq K-1$, and providing all the channel output observations $\yv^{n}_1,\yv^{n}_2, \cdots, \yv^{n}_{K}, y^{n}_0$ and the messages 
$W_1, \ldots,W_{L}$ to each of the remaining users (users $L+1,  \ldots, K$).  Then we derive the following upper bounds on the achievable
rates 
\begin{align}
  n \sum_{k=1}^{L} R_k &\le I(W_1, W_2, \cdots, W_{L}; \yv^{n}_1,\yv^{n}_2, \cdots, \yv^{n}_{L},    y^{n}_0 | \Omega^{n})  +  n \epsilon_n   ,  \label{eq:sumB1} \\
   n \sum_{k=L+1}^{K} R_k &\le I(W_{L+1}, \cdots, W_{K}; \yv^{n}_1, \ldots, \yv^{n}_K,y^{n}_0 | W_1, \ldots,W_{L},\Omega^{n})  +  n \epsilon_n  ,  \label{eq:sumB2} 
\end{align}
by applying Fano's inequality and some basic chain rules, and using the independence between the messages.
Then, we have
\begin{align}
  & n R_1 + nR_2 +\cdots +  nR_L  +  \frac{L}{K} \bigl (  n R_{L+1} + n R_{L+2} +   \cdots+   n R_K )    - n\bigl(1+\frac{L}{K}\bigr) \epsilon_n     \nonumber \\
  &\leq I(W_1, \cdots, W_{L}; \yv^{n}_1, \cdots, \yv^{n}_{L},    y^{n}_0 | \Omega^{n})    + \frac{L}{K} I(W_{L+1}, \cdots, W_{K}; \yv^{n}_1, \ldots, \yv^{n}_K,y^{n}_0 | W_1, \ldots,W_{L},\Omega^{n})    \nonumber \\
 & =  I(W_1,\cdots, W_{L}; \yv^{n}_1, \cdots, \yv^{n}_{L}| \Omega^{n}) + I(W_1,\cdots, W_{L};   y^{n}_0 | \yv^{n}_1, \cdots, \yv^{n}_{L}, \Omega^{n})   \nonumber \\ &\quad + \frac{L}{K} I(W_{L+1}, \cdots, W_K; \yv^{n}_1, \ldots, \yv^{n}_K | W_1, \ldots,W_{L},\Omega^{n}) \nonumber \\ &\quad + \frac{L}{K} I(W_{L+1}, \cdots, W_K; y^{n}_0 | \yv^{n}_1, \ldots, \yv^{n}_K, W_1, \ldots,W_{L},\Omega^{n})  \nonumber \\
& =  \underbrace{ h(\yv^{n}_1, \cdots, \yv^{n}_{L}| \Omega^{n})}_{\leq NLn\bigl(\log P+o(\log P)\bigr)}   - h(\yv^{n}_1, \cdots, \yv^{n}_{L}| W_1,\cdots, W_{L}, \Omega^{n})  \nonumber \\ &\quad + \frac{L}{K} h (\yv^{n}_1, \ldots, \yv^{n}_K | W_1, \ldots,W_{L},\Omega^{n}) - \frac{L}{K}  \underbrace{h( \yv^{n}_1, \ldots, \yv^{n}_K | W_1, \ldots,W_{K},\Omega^{n}) }_{ \geq n\cdot o(\log P) }  \nonumber \\ &\quad
+ \underbrace{ H( y^{n}_0 | \yv^{n}_1, \cdots, \yv^{n}_{L}, \Omega^{n})}_{ \leq  n\rM } - H( y^{n}_0 |  \yv^{n}_1, \cdots, \yv^{n}_{L}, W_1,\cdots, W_{L},\Omega^{n})   \nonumber \\ &\quad   +\underbrace{ \frac{L}{K} H( y^{n}_0 | \yv^{n}_1, \ldots, \yv^{n}_K, W_1, \ldots,W_{L},\Omega^{n}) }_{\leq H( y^{n}_0 | \yv^{n}_1, \ldots, \yv^{n}_{L}, W_1, \ldots,W_{L},\Omega^{n}) } - \frac{L}{K} \underbrace{ H(y^{n}_0 | \yv^{n}_1, \ldots, \yv^{n}_K, W_1, \ldots,W_{K},\Omega^{n})}_{\geq 0 } \nonumber\\
& \leq  NLn\bigl(\log P+o(\log P)\bigr) + n\rM  + n\cdot o(\log P) \nonumber \\ &\quad 
+ \frac{L}{K} h (\yv^{n}_1, \ldots, \yv^{n}_K | W_1, \ldots,W_{L},\Omega^{n})   - h(\yv^{n}_1, \cdots, \yv^{n}_{L}| W_1,\cdots, W_{L}, \Omega^{n})   \label{eq:K1up21742} \\
& \leq  NLn\bigl(\log P+o(\log P)\bigr) + n\rM  + n\cdot o(\log P) 
+ \frac{nL(1-\gamma)(K-L)}{K}  \bigl(N\log P+o(\log P)\bigr) ,  \label{eq:K1upfinal}
\end{align}
where \eqref{eq:K1up21742} follows from  $ h(\yv^{n}_1, \cdots, \yv^{n}_{L}| \Omega^{n}) \leq NLn\bigl(\log P+o(\log P)\bigr) $  and $H( y^{n}_0 | \yv^{n}_1, \cdots, \yv^{n}_{L}, \Omega^{n}) \leq  n\rM  $ and $h( \yv^{n}_1, \ldots, \yv^{n}_K | W_1, \ldots,W_{K},\Omega^{n})  \geq h( \yv^{n}_1, \ldots, \yv^{n}_K | \xv^n, W_1, \ldots,W_{K},\Omega^{n}) = h( \zv^{n}_1, \ldots, \zv^{n}_K)  = n\cdot o(\log P) $ and from  non-negativity of the entropy and the fact that conditioning reduces entropy.  The last inequality is from  Lemma~\ref{lm:DiffEntroKMIMO}.
Hence, dividing \eqref{eq:K1upfinal} by $n \log P$ and let $P\to\infty$, \eqref{eq:boundK1}
follows immediately for the case with $L<K$.

Considering the case with $L=K$, and starting from Fano's inequality,   we have
\begin{align}
& nR_1+nR_2+\cdots+nR_K  \nonumber\\
 &\leq I(W_1,\cdots,W_K; \yv^{n}_1, \cdots, \yv^{n}_K, y^{n}_0 |\Omega^{n})  +  n \epsilon_n  \nonumber\\
&=  I(W_1,\cdots,W_K; \yv^{n}_1, \cdots, \yv^{n}_K |\Omega^{n})  + I(W_1,\cdots,W_K; y^{n}_0 |\yv^{n}_1, \cdots, \yv^{n}_K, \Omega^{n})  + n \epsilon_n \nonumber\\
&= h(\yv^{n}_1, \cdots, \yv^{n}_K | \Omega^n) - h(\yv^{n}_1, \cdots, \yv^{n}_K | W_1,\cdots,W_K, \Omega^n) \nonumber\\ & \quad + H(y^{n}_0 |\yv^{n}_1, \cdots, \yv^{n}_K, \Omega^{n}) - H(y^{n}_0 |W_1,\cdots,W_K, \yv^{n}_1, \cdots, \yv^{n}_K, \Omega^{n}) +  n \epsilon_n \nonumber\\
&\le nKN\log P + n\rM - h(\yv^{n}_1, \cdots, \yv^{n}_K | W_1,\cdots,W_K, \Omega^n)  \nonumber\\ & \quad - H(y^{n}_0 |W_1,\cdots,W_K, \yv^{n}_1, \cdots, \yv^{n}_K, \Omega^{n}) +  n\cdot o(\log P) \label{eq:KMtmp785}\\
&\le nKN\log P + n \rM + n\cdot o(\log P) , \label{eq:KMtmp786}
\end{align}
where \eqref{eq:KMtmp785} follows from
$h(\yv^{n}_1, \cdots, \yv^{n}_K | \Omega^n)\le nKN \log P + n\cdot o(\log P)$ and  $H(y^{n}_0 |\yv^{n}_1, \cdots, \yv^{n}_K , \Omega^{n}) \le H(y^{n}_0) \le n \rM$; the last inequality follows from
the non-negativity of the entropy and the fact that
$h(\yv^{n}_1, \cdots, \yv^{n}_K  | W_1,\cdots,W_K, \Omega^n) \ge  n\cdot o(\log P)$.
Hence, dividing \eqref{eq:KMtmp786} by $n \log P$ and let $P\to\infty$, \eqref{eq:boundK1}
follows immediately for the case with $L=K$.




\end{document}